\documentstyle[preprint,prd,eqsecnum,aps,tighten,epsf]{revtex}
\begin{document}
\draft
\newcommand{\nl}{\nonumber \\}
\newcommand{\bea}{\begin{eqnarray}}
\newcommand{\eea}{\end{eqnarray}}
\def\vp{{\bf p}}
\def\ad{\overline D^0}
\def\ak{\overline K^0}
\def\ks{K^0_S}
\def\kl{K^0_L}
\def\ksp{K^{*+}}
\def\ksm{K^{*-}}
\def\kal{K_1(1270)}
\def\kah{K_1(1400)}
\def\dsp{D_s^+}
\def\dssp{D_s^{*+}}
\def\bcsp{B_c^{*+}}
\def\dil{\ell^+\ell^-}
\def\die{e^+e^-}
\def\dim{\mu^+\mu^-}
\def\ra{\rightarrow}
\def\dek#1{\times10^{#1}}
\newcommand{\bd}{B_d^0}
\newcommand{\bs}{B_s^0}
\newcommand{\be}{\begin{equation}}
\newcommand{\ee}{\end{equation}}
\newcommand{\bt}{\begin{table}}
\newcommand{\et}{\end{table}}
\newcommand{\btab}{\begin{tabular}}
\newcommand{\etab}{\end{tabular}}
\newcommand{\Vud}{\left|V_{ud}\right|}
\newcommand{\Vus}{\left|V_{us}\right|}
\newcommand{\Vcd}{\left|V_{cd}\right|}
\newcommand{\Vcs}{\left|V_{cs}\right|}
\def \zpc#1#2#3{Z.~Phys.~C {\bf#1}, #2 (19#3)}
\def \plb#1#2#3{Phys.~Lett.~B {\bf#1}, #2 (19#3)}
\def \prl#1#2#3{Phys.~Rev.~Lett.~{\bf #1}, #2 (19#3)}
\def \pr#1#2#3{Phys.~Rep.~{\bf #1}, #2 (19#3)}
\def \prd#1#2#3{Phys.~Rev.~D~{\bf#1}, #2 (19#3)}
\def \npb#1#2#3{Nucl.~Phys.~B {\bf#1}, #2 (19#3)}
\def \rmp#1#2#3{Rev.~Mod.~Phys.~{\bf#1}, #2 (19#3)}
\def \ea{{\it et al.}}
\def \ibid#1#2#3{{\it ibid.} {\bf#1}, #2 (19#3)}
\def \sjnp#1#2#3#4{Yad. Fiz. {\bf#1}, #2 (19#3) [Sov. J. Nucl. Phys.
{\bf#1}, #4 (19#3)]}

\title{
\flushright{\small\rm DOE/ER/40561-307-INT96-21-004}\\
\vskip 1in
{\bf Some implications of meson dominance in weak interactions}}
\author{Peter Lichard}
\address{
Department of Theoretical Physics,
Comenius University, 842-15 Bratislava, Slovak Republic\\
and Institute of Physics, Silesian University, 746-01 Opava,
Czech Republic\\
and Institute for Nuclear Theory, University of Washington, Seattle,
WA~98195
}
\maketitle
\begin{abstract}
The hypothesis is scrutinized that the weak interaction of hadronic
systems at low energies is dominated by the coupling of the
pseudoscalar, vector, and axial-vector mesons to the weak gauge bosons.
The strength of the weak coupling of the $\rho(770)$ meson is uniquely
determined by vector-meson dominance in electromagnetic interactions;
flavor and chiral symmetry breaking effects modify the coupling of
other vector mesons and axial-vector mesons. Many decay rates are
calculated and compared to experimental data and partly to predictions
of other models. A parameter-free description of the decay
$K^+\ra\pi^+\dil$ is obtained. Predictions for several not yet observed
decay rates and reaction cross sections are presented. The relation
between the conserved vector current hypothesis and meson dominance is
clarified. Phenomenological success of the meson dominance suggests
that in some calculations based on the standard model the weak
quark--antiquark annihilation and creation diagrams may be more
important than anticipated so far. The processes are identified where
the meson dominance fails, implying that they are governed, on the
quark level, by some other standard model diagrams.
\end{abstract}


\narrowtext

\section{INTRODUCTION}
\label{intro}

The idea of vector meson dominance (VMD), which was proposed a long
time ago \cite{vmdsource}, has proven to be very fruitful in describing
the electromagnetic interactions of hadrons at low energies. It is
routinely used, even today when the standard theory \cite{standard,gim}
provides a unified picture of all interactions among leptons and quarks.
The reason for the present day popularity of effective theories is the
difficulty encountered when building a bridge between the world of
quarks and gluons and that of hadrons.

According to the VMD hypothesis the electromagnetic interactions of
hadrons are mediated by neutral vector mesons ($\rho^0$, $\omega$,
$\phi$, and to a lesser extent also their higher recurrences) which
couple to the electromagnetic field $a_\mu$ according to the Lagrangian
\be
\label{vgamma}
{\cal L}_{\rm VMD}=-e\frac{m_\rho^2}{g_\rho}\bigg(V_{\rho^0}^\mu
+{1\over3}V_\omega^\mu
-\frac{\sqrt{2}}{3}V_\phi^\mu\bigg)a_\mu\ ,
\ee
where $V$'s are vector meson field operators. The presence
of the ratio of the $\rho$ mass squared to the $\rho\pi\pi$ coupling
constant $g_\rho$ ($g_\rho^2=36.56\pm0.29$) is required by the
normalization condition $F_\pi(0)=1$ for the pion form factor. The
other factors follow from invariance under the $U$-spin $SU(2)$
subgroup of the (flavor) $SU(3)$ group and the assumption that the
physical $\omega$ meson does not contain $s$-quarks.

The idea of the universality of the vector current led very soon
to the application of VMD in weak interactions. The early development
was reviewed in \cite{chounet,bardin}. But, unlike in QED, the
transition amplitudes between the weak gauge bosons and mesons are
nonvanishing also for pseudoscalar and axial-vector mesons. It is
therefore natural to generalize the VMD and assume that the weak
interaction of hadronic systems is dominated by the coupling of
pseudoscalar, vector, and axial-vector mesons to the gauge bosons
$W^\pm$ and $Z^0$. We will refer to this hypothesis as meson dominance
(MD).

Qualitative support for MD in weak interactions comes at least from
two sources: (1) Individual pseudoscalar, vector and axial-vector mesons
are copiously produced in the decay of the tau lepton; (2) MD naturally
explains why the ratios among various charge configurations of hadronic
final states in weak decays often follow the rules implied by isospin
invariance, which is otherwise violated in weak interactions.

The MD hypothesis has two components. Firstly, the assumption that
the weak interaction of hadronic systems is dominated by the coupling
of individual mesons to the weak gauge bosons means, on the quark
level, restriction to a certain class of perturbative expansion
diagrams. This class does not include, e.g., the penguin and box
diagrams. Secondly, in order to make the MD a quantitative concept
we have to establish the effective Lagrangian of the interaction
between mesons and gauge bosons. This will be done in Sec.~\ref{deriv}.

The question arises whether we really need a simple and approximate
phenomenological approach to the electroweak interaction since we
believe that the fundamental theory exists, which allows one to
calculate everything from first principles. We think that the reasons
for exploring MD are twofold.

It is true that the basic electroweak diagrams of most decay modes
are relatively simple. But, as a matter of fact, the QCD effects play
an important role. The calculation of QCD corrections to the basic
electroweak diagrams is the most difficult and involved part
\cite{buchalla}. In contrary, the MD approach takes advantage of the
fact that the Mother Nature made some QCD calculations for us, even
nonperturbatively, when she built hadrons. It is not true in general.
The results of some QCD calculations (e.g., QCD penguin diagrams, QCD
corrections to the weak and electromagnetic penguins and to the box
diagrams) are not accessible so easily and MD cannot be explored in such
cases. Anyhow, an approximately correct description of some process by
the MD may have heuristic value for the more fundamental approaches by
showing which quark diagrams may be most important. We will try
to illustrate this point in Sec.~\ref{neutral}.

Another reason for our considering the MD approach is the relations
between theory and experiment. The fundamental theories or models more
sophisticated than MD do not often provide simple formulas for various
distributions that would be suitable for use by experimentalists to fit
their data. As a consequence, formulas lacking dynamic motivation or
sometimes even violating the basic principles of quantum mechanics are
used. The MD approach may be able to offer formulas, even if
approximate, that are simple and reflect, at least in a crude way,
the underlying dynamics.

We will try to keep this paper selfcontained and provide all the
formulas and numbers we used. The values of input parameters (masses,
hadronic widths, lifetimes) needed in our calculations were taken from
\cite{pdg}. Unless stated otherwise, so were the experimental values
of branching fractions to which we compare our results. The quoted
errors of our results reflect only those of input parameters. No attempt
has been made to assess the systematic uncertainties of the meson
dominance approach. Not to expand the scope of the paper too much,
we do not compare, with a few exceptions, our results to those of
existing models. The references to the latter can be traced back from
the most topical ones \cite{isgw2,du96}, some of them are mentioned
in this paper later on.

Throughout the paper $P$, $S$, $V$, and $A$ will be used as generic labels
for pseudoscalar, scalar, vector, and axial-vector mesons, respectively.
In Lagrangians, the field operators will always be those of individual
members of isospin multiplets. The isospin symmetry of hadronic
Lagrangians will be ensured through the relations among the coupling
constants for different charge combinations of participating mesons.

In the next section, we write effective Lagrangians describing the
coupling of vector and axial-vector mesons to charged and neutral
weak gauge bosons $W^\pm$ and $Z^0$ and define the parameters of our
approach. Section~\ref{taudecay} deals with the decays of the
$\tau$ lepton, some used as a source of information about
the MD parameters.
In Sec.~\ref{psps}, we investigate vector current processes in
which the core of hadronic part is the $PPV$ vertex, with vector
meson $V$ converting into the charged gauge boson. They include
the following types of decays: $P_1\ra P_2+\ell\nu_\ell$,
$P_1\ra P_2+P_3$, $P_1\ra P_2+V$, and $P_1\ra P_2+A$. We calculate
also the cross sections of the antineutrino-electron and meson-electron
binary reactions that are related to the semileptonic decay shown above
by crossing symmetry.
In Section~\ref{neutral} we show that MD leads to a parameter-free
formula for the rate of the decay $K^+\ra\pi^+\die$ that agrees
with the experimental value. Predictions for the dimuon mode and
for the transitions of $D^+$ and $D_s$ mesons into the same final
state are also made.
Section~\ref{cvcvermd} is devoted to the relation between the
conserved vector current (CVC) and MD hypotheses.
We summarize our main points and add a few
comments in Sec.~\ref{comments}.
Some related issues are deffered to the Appendices. In Appendix
\ref{detccvvp} we extract some hadronic coupling constants from
data on hadronic and radiative decay widths.
Appendix \ref{genff} shows the decay rate of
$P_1\ra P_2+\ell\nu_\ell$ for arbitrary form factors, which was used
in Sec.~\ref{cvcvermd}.

\section{Defining meson dominance in weak interactions}
\label{deriv}

In this section we will use plausible arguments based on the standard
model Lagrangian and the VMD in electromagnetic sector in order to
find the effective Lagrangian that describes the coupling of vector
and axial-vector mesons to the weak gauge bosons, both charged and
neutral. In a search for it we first discuss the dynamical content
of the VMD in electromagnetic interactions from the quark model point
of view. Then we will apply the same procedure to the weak interactions.

\subsection{Vector meson dominance in electromagnetic interactions}
\label{vmd}

The electromagnetic part of the standard model Lagrangian
\be
\label{elmaglag}
{\cal L}_{\rm EM}(x)=-j^\mu(x) a_\mu(x)
\ee
contains the electromagnetic field operator $a_\mu$ and the quark
electromagnetic current
\[
j^\mu(x)=e\sum_{i=1}^3 \left[\ {2\over3}\ \bar u_i(x)\gamma^\mu u_i(x)
-{1\over3}\ \bar d_i(x)\gamma^\mu d_i(x)\right] ,
\]
where $u_i$ ($d_i$) denotes the field operator of the up (down) quark
from the $i$th generation. The matrix element of an electromagnetic
process with a $\rho^0$ in the initial state will contain the factor
\be
\label{mej0}
\langle0|j^\mu(0)|p,\lambda\rangle_{\rho^0}
={2\over3}e
\ \langle0|\bar u(0)\gamma^\mu u(0)|p,\lambda\rangle_{\rho^0}
-{1\over3}e
\ \langle0|\bar d(0)\gamma^\mu d(0)|p,\lambda\rangle_{\rho^0}
\ ,
\ee
where $p$ and $\lambda$ are the four-momentum and polarization of $\rho^0$,
respectively. Only quarks of the first generation matter. The matrix
elements must transform like four-vectors. The only two four-vectors
we have at our disposal are the four-momentum $p^\mu$ of the $\rho^0$ and its
polarization vector $\epsilon^\mu(p,\lambda)$. Because we are interested
in low energy interactions, we will neglect the term proportional
to the four-momentum. We thus write
\be
\label{frho}
\langle0|\bar u(0)\gamma^\mu u(0)|p,\lambda\rangle_{\rho^0}
=F_\rho\epsilon^\mu(p,\lambda)\ ,
\ee
where $F_\rho$ is a constant. Isospin invariance together with
the isovector character of the $\rho$ implies that
\[
\langle0|\bar d(0)\gamma^\mu d(0)|p,\lambda\rangle_{\rho^0}
=-\langle0|\bar u(0)\gamma^\mu u(0)|p,\lambda\rangle_{\rho^0}\ .
\]
Putting it into (\ref{mej0}) and using Eq.~(\ref{frho}) we get
\be
\label{matrix4}
\langle0|j^\mu(0)|p,\lambda\rangle_{\rho^0}
=eF_\rho\epsilon^\mu(p,\lambda)
\ .
\ee
On the other hand, we have the relation
\be
\label{matrix5}
\langle0|V_{\rho^0}^\mu(0)|p,\lambda\rangle_{\rho^0}
={\cal N}\epsilon^\mu(p,\lambda)
\ .
\ee
Both $F_\rho$ and the constant ${\cal N}$ depend on the normalization of
one-particle states,
but their ratio does not. Comparing (\ref{matrix4}) and (\ref{matrix5})
we see that the Lagrangian obtained from (\ref{elmaglag}) by the substitution
\[
j^\mu(x)\ra e\frac{F_\rho}{{\cal N}}V_{\rho^0}^\mu(x)
\]
gives the same low energy matrix element as the original Lagrangian
(\ref{elmaglag}).
In order the get correctly the first term in Eq. (\ref{vgamma}), which
is fixed by the normalization of the pion form factor, the following
relation must be held
\be
\label{fovern}
\frac{F_\rho}{\cal N}=\frac{m_\rho^2}{g_\rho}
\ee

Repeating all the steps for the isoscalar $\omega$ and $\phi$ mesons,
and using $SU(3)$ invariance relations
\be
\label{omphi}
\langle0|{\overline u}(0)\gamma^\mu u(0)|p,\lambda\rangle_{\omega}
=\frac{1}{\sqrt{2}}
\langle0|{\overline s}(0)\gamma^\mu s(0)|p,\lambda\rangle_{\phi}
=\langle0|{\overline u}(0)\gamma^\mu u(0)|p,\lambda\rangle_{\rho^0}\ ,
\ee
together with (\ref{fovern}), we obtain the other terms of
Eq.~(\ref{vgamma}). We assumed that the $\phi$ meson transforms like a
pure $\bar{s}s$ state.\footnote{Relations (\ref{omphi}) and other
of this kind  stem from the transformation properties of
the wave functions and field operators, and do not mean that we
ignore the gluon or sea quark content of the meson wave functions.}

We consider it important to stress that $g_\rho$ in Eq.~(\ref{vgamma})
and, as a consequence, also in the weak Lagrangians we are going to
introduce in what follows, is the coupling constant of the
$\rho\pi\pi$ interaction, determined from the $\rho\ra\pi\pi$ decay
width. This is strictly required by the condition $F_\pi(0)=1$ on the
electromagnetic form factor of the $\pi^+$ meson calculated from
(\ref{vgamma}). In some papers this important constrain was ignored
and the $g_\rho$ was determined from the electronic decay width of the
$\rho^0$ meson. It is true that if one calculates the rate of the
$\rho^0\ra e^+e^-$ decay from the Lagrangian (\ref{vgamma}) with
$g_\rho$ determined from $\rho\ra\pi\pi$, the result is obtained which
is smaller by a factor of $1.45\pm0.07$ than the experimental value.
But this is, as we will show elsewhere, the effect of higher $\rho$
resonances. It is improper to mimic this effect  by violating the
normalization condition of the form factor.

Also, if higher mesons from the $\rho$ family are added into
(\ref{vgamma}), the coupling of $\rho(770)$ to the electromagnetic
field may be modified. The normalization condition for the pion
form factor implies a definite relation among $\rho\gamma$ coupling
constants. The safest way of accounting for the influence of higher
$\rho$ resonances is to replace the properly normalized form factor
induced by $\rho(770)$ with a properly normalized form factor containing
all considered resonances. Preference is for the experimentally
determined ones, if available.

\subsection{Meson dominance in weak interactions}
\label{md}

Now we are going to apply the same procedure to the weak interactions.
First we will fix the coupling of the charged $\rho$ mesons to the
charged weak gauge bosons $W^\pm$. We start again from
the standard model Lagrangian, this time from the part that
exhibits the charged current weak interaction of the $u$ and $d$ quarks,
\begin{eqnarray}
\label{lagud}
{\cal L}_{ud}&=&-W^-_\mu j_{ud}^\mu +{\rm h.c.}\ , \nl
j_{ud}^\mu&=&\frac{g}{2\sqrt{2}}V_{ud}\ {\overline d}\gamma^\mu
(1-\gamma_5)u \ ,
\end{eqnarray}
where $g=e/\sin \theta_W$ is the electroweak coupling constant and
$V_{ud}$ is the relevant element of the Cabibbo-Kobayashi-Maskawa (CKM)
matrix \cite{cabibbo,kobmask}.
The matrix element for a process with a $\rho^+$ in the initial state
is proportional to
\[
\langle0|j^\mu_{ud}(0)|p,\lambda\rangle_{\rho^+}=\frac{g}{2\sqrt{2}}V_{ud}
\langle0|{\overline d}(0)\gamma^\mu (1-\gamma_5)u(0)|p,\lambda\rangle_{\rho^+}
\ .
\]
The axial-vector part does not contribute, and for the vector part
we can write
\be
\label{merhopl}
\langle0|\bar{d}(0)\gamma^\mu u(0)|p,\lambda\rangle_{\rho^+}
=\sqrt{2}
\ \langle0|\bar{u}(0)\gamma^\mu u(0)|p,\lambda\rangle_{\rho^0}
=\sqrt{2}\ F_\rho\epsilon^\mu(p,\lambda)\ .
\ee
Writing  an equation analogous to (\ref{matrix5}), and using the value
of $F_\rho/{\cal N}$ as implied by the VMD for electromagnetic
interactions, see (\ref{fovern}), we come to the conclusion that the
effective Lagrangian
for the low energy weak interaction of the charged $\rho$ mesons is
\[
{\cal L}_{\rho^\pm}=-\frac{gm_\rho^2}{2g_\rho}V_{ud}\ W_\mu^-
V_{\rho^+}^\mu\ +{\rm h.c.}
\]

Now let us investigate the coupling of the $K^{*+}$ to $W^+$.
The corresponding piece of the standard model Lagrangian is
\bea
\label{lagus}
{\cal L}_{us}&=&-W^-_\mu j_{us}^\mu +{\rm h.c.}\ , \nl
j_{us}^\mu&=&\frac{g}{2\sqrt{2}}V_{us}\ {\overline s}\gamma^\mu
(1-\gamma_5)u \ .
\eea
In analogy with (\ref{merhopl}) we define the constant $F_{K^*}$ by
the relation
\be
\label{mekstpl}
\langle0|\bar{s}(0)\gamma^\mu u(0)|p,\lambda\rangle_{\ksp}
=\sqrt{2}\ F_{K^*}\epsilon^\mu(p,\lambda)\ .
\ee
Equation (\ref{matrix5}) is valid also for $\ksp$ with the same
value of the
normalization constant ${\cal N}$. Using (\ref{fovern}) and defining
\be
\label{wkst}
w_{K^*}=\frac{F_{K^*}}{F_\rho}\frac{m_\rho^2}{m_{K^{*+}}^2}
\ee
we arrive at the conclusion that the Lagrangian
\[
{\cal L}_{K^{*\pm}}=-\frac{gm_{\ksp}^2}{2g_\rho}\ w_{K^*}V_{us}
\ W_\mu^-V_{\ksp}^\mu\ +{\rm h.c.}
\]
gives the same values of all observables in low energy processes
with a $K^{*\pm}$ in
the initial or final state as the standard model Lagrangian
(\ref{lagus}) after the constant $w_{K^*}$ is properly adjusted.
In the $SU_f(3)$ symmetry was exact we would have $w_{K^*}=1$.

The case of the axial-vector meson $a_1$ can be handled in the same way.
We define the constant $F_{a_1}$ by the relation
\be
\label{fa1}
\langle0|{\overline u}(0)\gamma^\mu\gamma_5 u(0)|p,\lambda\rangle_{a_1^0}
=F_{a_1}\epsilon^\mu(p,\lambda)\ .
\ee
With Eq.~(\ref{fovern}) in mind we further define
\be
\label{wa1}
w_{a_1}=\frac{F_{a_1}}{F_\rho}\frac{m_\rho^2}{m_{a_1}^2}\ .
\ee
The Lagrangian
\[
{\cal L}_{a_1^\pm}=\frac{gm_{a_1}^2}{2g_\rho}w_{a_1}V_{ud}\ W_\mu^-
A_{a_1^+}^{\mu}\ +{\rm h.c.}
\]
then leads to the same matrix elements for processes with the $a_1$ in
the initial or final state as the original standard model Lagrangian
(\ref{lagud}). In that sense it represents an effective Lagrangian for
the weak interaction of $a_1^\pm$ mesons. The constant $w_{a_1}$ is
a phenomenological parameter of the MD approach and should be determined
from data. In the chiral limit $m_u=m_d=0$ the $\rho$ and $a_1$ mesons
constitute a parity degenerate doublet and the $w_{a_1}$ would be unity.

The $a_1$ meson belongs to the $^3P_1$ octet of axial-vector mesons.
There has not been any experimental indication that its counterpart
from the $^1P_1$ octet, namely, the $b_1$ axial-vector meson, also
couples to the weak gauge bosons. For instance, it has not been
identified among the decay products of the $\tau$ lepton. One possible
explanation follows. The valence quark and antiquark in the $b_1$ form
a singlet spin state. Their helicities in the meson rest frame thus tend
to be equal, what leads to a small matrix element of the weak quark
current. But the coupling of the $b_1$ to the gauge bosons, in other
words, the existence of a second class axial-vector current
\cite{class}, is not ruled out absolutely by the quark model approach.
There is no obvious reason for the matrix element of the type
(\ref{fa1}) written for the $b_1$ meson to vanish identically.

To complete our considerations of the MD in the charged current weak
interactions let us recall
that quarks from higher generations enter the standard model Lagrangian
in the same way as the $u$ and $d^\prime$ quarks
from the first generation. This suggests that the most general form of the
charged current MD is
\be
\label{ccvavd}
{\cal L}_{c.c.}=-\frac{g}{2g_\rho}\ W_\mu^-
\bigg[\sum_{V=\rho^+,\ksp,\cdots}
m_V^2V_Vw_VV^\mu-\sum_{A=a_1^+,K_1^+,\cdots}m_A^2V_Aw_AA^{\mu}\bigg]
\ +\ {\rm h.c.}\
\ee
Here $V^\mu$  and $A^\mu$ are the field operators of positively charged
vector ( $\rho^+$, $K^{*+}$, $D^{*+}$,
$D^{*+}_s$, $B^{*+}$) and axial-vector mesons [$a_1^+$, $K_1(1400)^+$,
$D_1^+$, $D_{s1}^+$], respectively.
Unlisted states either do not exist or have not yet been discovered.
$V_V$ and $V_A$ are the elements of the CKM matrix that correspond
to the valence quark composition of the particular vector or
axial-vector meson. To make Eq. (\ref{ccvavd}) compact we have
introduced $w_{\rho^+}\equiv 1$.
In the case of exact $SU_f(6)$ symmetry also the other $w_V$ would be
equal to one and all $w_A$ would be equal to $w_{a_1^+}$.
The actual values may be different.

Anyhow, the $w_A$ should not differ too much from the $w_V$
of the vector mesons with the same flavor, because the
corresponding vector and axial-vector mesons form a chiral symmetry
doublet. The relative sign of the vector and
axial-vector parts of (\ref{ccvavd}) is important for processes
to which they both contribute.

The parameters $w_V$ and $w_A$ enter formulas for observable quantities
in combinations with other parameters (CKM matrix elements, strong
interaction coupling constants). In some cases it makes their extraction
from the data impossible. This does not diminish appreciably the
predictive power of the MD because the same product of parameters
determines the rates or cross sections of several processes. We can thus
fix the normalization using one piece of data and other quantities then
are predicted by the MD. This approach will be used extensively in
Section~\ref{psps}.

In the cases when the vector or axial-vector meson that couples to
the gauge boson appears as one of the final state particles it is
useful to define the quantities
\be
\label{ym}
Y_M=\left|w_MV_M\right|^2 ,
\ee
where $M$ stands for any of the charged vector and axial-vector mesons.
$V_M$ is the element of the CKM matrix pertinent to the valence quark
and antiquark of the particular meson, and $w_M$ is the parameter in
the effective charged current MD Lagrangian (\ref{ccvavd}) that appears
in the $M^+W^+$ junction. The $Y_M$'s that will be fixed by data
later on are shown in Table~\ref{ymtab}.

A few more words are needed about the strange axial-vector mesons that
exist in two sorts, $\kal$ and $\kah$. Analysis of the
branching fractions of the $\tau^-$ lepton suggests that the coupling
of $\kah$ to the weak gauge bosons is stronger than that
of $\kal$. First of all, the $\tau^-$ branching fraction to $\kah^-$
is $(8\pm4)\dek{-3}$, to $\kal^-$, $(4\pm4)\dek{-3}$. Next,
the branching ratios of $\tau^-$ to $K^-\pi^+\pi^-$, $\ak\pi^-\pi^0$,
and $K^-\pi^0\pi^0$ systems are compatible with 4:4:1 ratio, which
is typical for decay of an $I=1/2$ resonance ($K_1^-$) to those systems
through the $(K^*\pi)^-$ intermediate state. It again points to
$\kah$ with its branching fraction to $K^*\pi$ of $(94\pm6)\%$
rather than to $\kal$ [$(16\pm5)\%$]. Finally, if the axial strange
mesons that couple to the $W$ gauge boson were the $^3P_1$ state
$K_{1A}$ or the $^1P_1$ state $K_{1B}$, nearly equal mixes of the $\kal$
and $\kah$, then the $\kal$ and its decay products would be more visible
in $\tau$ decays. Also, in the recent work \cite{beldjoudi} current
algebra was applied in the three-pseudoscalar-meson decays of the $\tau$
lepton. The $K^-\rho^0\nu_\tau$ mode [the dominant strong decay mode of
$K_1(1270)$] was shown to be consistent with zero.

The higher vector and axial-vector recurrences are not explicitly shown
in (\ref{ccvavd}). Generally, their influence will be difficult to take
into account due to insufficient knowledge of their couplings to other
hadrons. In some cases (certainly in the phenomenologically most
important case of the $\rho$ meson family) they can be taken into
account by replacing a simple pole contribution, stemming from the
virtual meson propagator, by an empirically determined electromagnetic
form factor. When appropriate, we will use that of \cite{dubnicka}.

When using the same procedure to determine the coupling of the vector
and axial-vector mesons to the $Z$ gauge boson, we find that
only the truly neutral (all additive quantum numbers vanishing)
mesons can couple. The coupling of those consisting of a down quark and
a down antiquark from different generations
is proportional to the off-diagonal elements of the product
$V^\dagger V$ and is therefore forbidden by the unitarity of the
CKM matrix $V$. The neutral mesons formed of valence up quark and
antiquark from different generations
are excluded by the form of the neutral weak current itself.
The effective Lagrangian describing the interaction of truly neutral
vector and axial-vector mesons with the Z boson has the form
\be
\label{ncvavd}
{\cal L}_{\rm n.c.}=-\frac{g}{2g_\rho\cos\theta_W}
Z_\mu\bigg[\sum_{V=\rho^0,\omega,\cdots} m_V^2G_VV^\mu\ -
\sum_{A=a_1^0,f_1,\cdots} m_A^2G_AA^\mu\bigg]\ ,
\ee
where $G_{\rho^0}=w_{\rho^0}\ (1-2\sin^2\theta_W)$, $G_{\omega}=-
2/3\ w_\omega\ \sin^2\theta_W$,
$G_{J/\psi}=\sqrt{2}\ w_{J/\psi}\ (1/2-4/3\ \sin^2\theta_W)$, and
$G_{V}=\sqrt{2}\ w_V\ (-1/2+2/3\ \sin^2\theta_W)$ for $V=\phi,\Upsilon$.
For axial-vector mesons we have $G_{a_1^0}=w_{a_1^0}$,
$G_{A}=-w_A/\sqrt{2}$ for the pure $\bar{d}d$ states
[$A=f_1(1510),\ \chi_{b1}$],
$G_{A}=w_A/\sqrt{2}$ for the pure $\bar{u}u$ states ($\chi_{c1}$), and
$G_{A}=0$ for $f_1(1285)$,
the isoscalar axial-vector counterpart of the $\omega(782)$.
Most of the constants $w_V$ and $w_A$ represent new
parameters, with values expected not to be far from unity. Isospin
symmetry enables to relate some of them to the corresponding parameters
of the charged current Lagrangian (\ref{ccvavd}), e.g., $w_{\rho^0}=
w_\omega=w_{\rho^+}\equiv 1$, $w_{a_1^0}= w_{a_1^+}$.
Under exact $SU_f(3)$ symmetry we would also have $w_\phi=1$.

The weak interaction of pseudoscalar mesons is routinely described
by the Lagrangian
\be
\label{pslag}
{\cal L}_P=-i\ \frac{g}{2\sqrt{2}}\ W^-_\mu\sum_{P=\pi^+,K^+,\cdots}
f_P\ V_P\ \partial^\mu\varphi_P\ +\ {\rm h.c.}
\ee
where $V_P$ is the element of the CKM matrix pertinent to valence
quark and antiquark of the meson $P$ and $f_P$ is the
pseudoscalar-meson decay constant defined for the $\pi^+$ meson by
\be
\label{fpi}
\langle 0|\bar{d}(0)\gamma^\mu\gamma_5u(0)|p\rangle_{\pi^+}=
i\ f_{\pi^+}\ p^\mu
\ee
and analogously for other mesons. Observables (decay rates, cross
sections) of the processes with the pseudoscalar meson $P$ either
in the initial or final state will be proportional to the
quantity\footnote{In this paper we will not be faced with the necessity
to consider the interference of several diagrams.}
\be
\label{z}
Z_P=\left|f_PV_P\right|^2.
\ee
The values of these parameters for different pseudoscalar mesons can be
determined from their leptonic branching fractions $P\ra\ell\nu$
and are shown in Table~\ref{fvtab}. In the case of the $\pi^+$ and $K^+$
leptonic decays the radiative corrections are important. We used the
prescription defined in Suzuki's article in \cite{pdg}, p. 319. For the
$D^+$ meson only an upper limit on the leptonic branching fraction is
known experimentally. Here we used the recent lattice calculation
\cite{soni} result $f_{D^+}=(208\pm35\pm12)$~MeV and
$\Vcd=0.224\pm0.016$ from \cite{pdg}. We summed the errors quadratically.

From the MD point of view the coupling of scalar mesons to weak gauge
bosons is not excluded. However, the success of the conserved vector
current (CVC) hypothesis shows that this coupling, which represents a
second-class vector current \cite{class}, must be negligible.
Nevertheless, let us define the scalar-meson decay constant of the
$a_0^+$ meson analogously to (\ref{fpi}) by means of the matrix
element of the vector part of the weak current
\be
\label{fa0}
\langle 0|\bar{d}(0)\gamma^\mu 0u(0)|p\rangle_{a_0^+}=
i\ f_{a_0^+}\ p^\mu.
\ee
and similarly for other charged scalar mesons [in fact, apparently
only one exists--$K_0(1430)$]. The effective Lagrangian is given as
\be
\label{slag}
{\cal L}_S=i\ \frac{g}{2\sqrt{2}}\ W^-_\mu\sum_{S=a_0^+,\ K_0^+}
f_S\ V_S\ \partial^\mu\varphi_S\ +\ {\rm h.c.}\ .
\ee
In the next section we will show on the basis of experimental data
that the decay constant of the $a_0^+$ meson
is at least twenty times smaller than that of the $\pi^+$.

In this paper we will not consider processes in which a truly neutral
spin zero meson couples to the neutral weak gauge boson $Z$. We
therefore do not write the corresponding Lagrangians here.

What we have done in this section can only be considered as a more
or less educated guess, not derivation, of what the effective
Lagrangians for the weak interaction of vector, axial-vector,
pseudoscalar, and scalar vectors may look like. Moreover, we have
so far considered only real (incoming or outgoing) mesons. Going off
mass-shell may, in principle, convert the Lagrangian parameters $w_V$,
$w_A$, $f_P$, and $f_S$ into arbitrary functions of meson virtualities
$p^2$. To proceed further we will neglect this possibility and postulate
the validity of Lagrangians as given in Eqs.~(\ref{ccvavd}),
(\ref{ncvavd}), (\ref{pslag}), and (\ref{slag}) for both real and
virtual mesons. This postulate, together with the restriction to a
certain class of quark diagrams, discussed in Section~\ref{intro},
constitute the main ingredients of the meson dominance in weak
interactions.

\section{MESON DOMINANCE AND DECAY MODES OF THE \protect{$\tau$} LEPTON}
\label{taudecay}

The decays of the $\tau$ lepton have intensively been studied both
experimentally and theoretically. Theoretical methods range from VMD to
chiral perturbation theory, see, e.g.,
\cite{decker96,fink96,davoud,colangelo,li} and references therein.
The main aim of this section is to extract some MD parameters that will
be used in Secs.~\ref{psps} and \ref{neutral} in the calculations
of decay rates of pseudoscalar mesons and cross sections of reactions
involving them. Nevertheless, in order to assess the possibilities and
limitations of the MD approach we consider it useful to show its
predictions for the $\tau$ lepton decay modes, even if many of them have
already been obtained by other authors. In some cases the MD works well,
even for such a complex decay mode as $\tau^-\ra\pi^-\pi^0\eta\nu_\tau$.

Some formulas presented here were derived by Tsai \cite{tsai} for
decays of heavy leptons before the $\tau$ lepton was actually discovered
\cite{perl}. His theoretical input included lepton universality, conserved
vector current (CVC) hypothesis \cite{cvc}, VMD in electromagnetic
interactions, and the Weinberg \cite{weinbsum} and  Das-Mathur-Okubo
\cite{dmo} sum rules.

\begin{figure}
\begin{center}
\leavevmode
\setlength \epsfxsize{15cm}
\epsffile{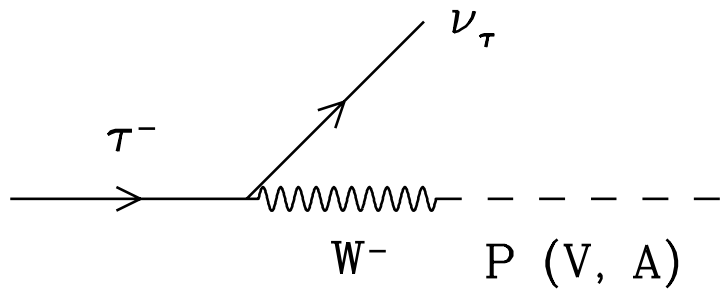}
\end{center}
\caption{Matrix element of the $\tau^-$ decay to neutrino and a
pseudoscalar, vector, or axial-vector meson.}
\label{taurhof}
\end{figure}

The formula for the partial decay width of $\tau^-$ into a pseudoscalar
meson and neutrino (see Fig.~\ref{taurhof}) is well known and can be
found, e.g., in \cite{tsai,okun,pietschm}. For the reader's convenience
and later reference it is shown also here.
\be
\label{taups}
\Gamma_{\tau^-\ra P^-\nu_\tau}=\frac{G_F^2 m_{\tau^-}^3 Z_P}{16\pi}
\left(1-\frac{m_P^2}{m_{\tau^-}^2}\right)^2.
\ee
Using the values of $Z_P$ parameters as given in Table~\ref{fvtab}
and the mean lifetime of the $\tau^-$ we get the branching fractions of
the $\pi^-$ and $K^-$ mode shown in Table~\ref{tautabl} together with
results of the evaluation of other $\tau^-$ decay modes described below.

The Feynman diagram that corresponds to the decay of the $\tau$ lepton
to a $\rho$ meson and a neutrino is shown in Fig.~\ref{taurhof}.
Using the MD Lagrangian (\ref{ccvavd}) it is easy to write down the
corresponding matrix element. The resulting formula for the partial
decay width, first derived  by Tsai \cite{tsai}, is
\be
\label{taurho}
\Gamma_{\tau^-\ra\nu_\tau\rho^-}=
\left(\frac{G_F\Vud}{g_\rho}
\right)^2\frac{m^2_\rho}{8\pi m^3_\tau}\left(m_\tau^2-m_\rho^2\right)^2
\left(m_\tau^2+2m_\rho^2\right)\ .
\ee
Numerically, Eq. (\ref{taurho}) yields a branching fraction of 19.0\%.

\begin{figure}
\begin{center}
\leavevmode
\setlength \epsfxsize{15cm}
\epsffile{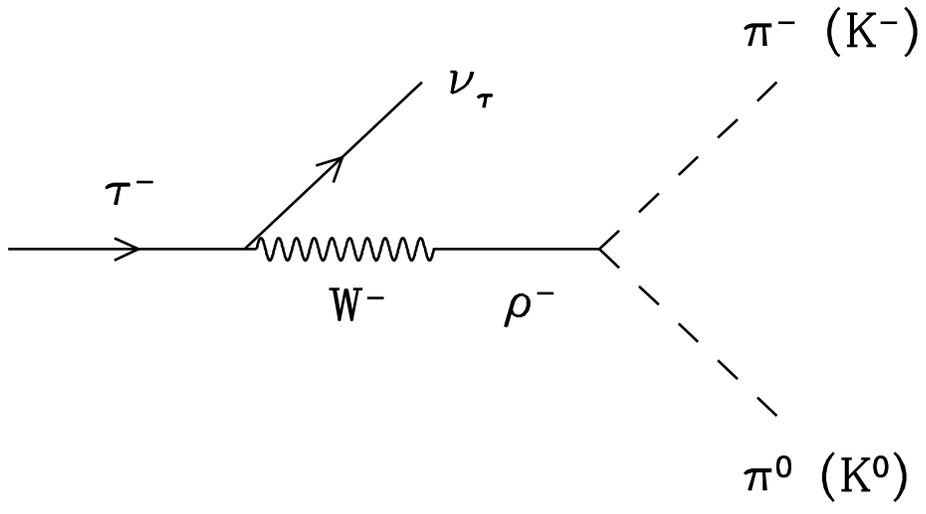}
\end{center}
\caption{Matrix element of the decays
$\tau^-\ra\pi^-\pi^0\nu_\tau$ and
$\tau^-\ra K^- K^0\nu_\tau$.}
\label{taupipif}
\end{figure}

To correct for the finite $\rho$ width, we will consider the three-body
decay $\tau^-\ra\pi^-\pi^0\nu_\tau$ going via $\rho^-$ in intermediate
state, see Fig.~\ref{taupipif}. Let us recall first that if the decaying
particle possesses spin zero or if we average over its spin states, then
the usual formula for the three-body decay $a\ra 1+ 2 +3$ (see, e.g.,
\cite{pdg}, p. 176) simplifies to
\be
\label{3body}
d\Gamma=\frac{1}{8(2\pi)^4m_a^2}\overline{\left|{\cal M}\right|^2}
\left|\vp_1\right|\left|\vp^*_2\right|\ dM_{23}\ d\Omega_2^*\ ,
\ee
where $\vp_1$ is the momentum of particle 1 in the rest frame of the
decaying particle, $\vp_2^*$ is the momentum of particle 2 in the rest
frame of 2 and 3, $d\Omega_2^*$ is the corresponding solid angle element,
and $M_{23}$ is the mass of the 2-3 subsystem. The bar over the matrix
element squared signifies the sum over the final and average over the
initial states.

The interaction among a vector field and two pseudoscalar fields
is described by the Lagrangian
\be
\label{lagvpsps}
{\cal L}_{VP_1P_2}=ig_{VP_1P_2}V_\mu\ \varphi_1
\stackrel{\longleftrightarrow}{\partial^\mu}
\varphi_2\ +{\rm h.c.}
\ee
If the decay $V\ra P_1+P_2$ is kinematically allowed then its rate
comes out from (\ref{lagvpsps}) as
\be
\label{vp1p2}
\Gamma_{V\ra P_1+P_2}=\frac{g^2_{VP_1P_2}}{48\pi m_V^5}
\lambda^{3/2}(m_V^2,m_{P_1}^2,m_{P_2}^2)\ ,
\ee
where
\be
\label{triangle}
\lambda(x,y,z)=x^2+y^2+z^2-2xy-2xz-2yz \
\ee
is the so called triangle function.
In the $\rho\pi\pi$ case the coupling constants for all three charge
combinations have the same absolute value $g_\rho$.

Using Lagrangians (\ref{ccvavd}) and (\ref{lagvpsps}), the
three-body decay formula (\ref{3body}) and neglecting the difference
between the $\pi^-$ and $\pi^0$ masses we arrive at the partial width
per unit interval in the $\pi^-\pi^0$ system mass
\be
\label{taupipi}
\frac{d\Gamma_{\tau^-\ra\pi^-\pi^0\nu_\tau}}{dM}=\frac{(G_F\Vud)^2}
{192\pi^3m_\tau^3s}
\left(m_\tau^2-s\right)^2
\left(m_\tau^2+2s\right)\left(s-4m_\pi^2\right)^{3/2}
\left|F(s)\right|^2  ,
\ee
where $s=M^2$ and
\be
\label{fmsq}
\left|F(s)\right|^2=\frac{m_\rho^4}{\left(s-m_\rho^2\right)^2
+m_\rho^2\Gamma_\rho^2}\ .
\ee
Let us note that $g_\rho$ coming from the Lagrangian (\ref{ccvavd})
canceled with that from the $\rho^-\pi^-\pi^0$ vertex.
To account for contributions from higher $\rho$ resonances, in actual
calculations we replaced (\ref{fmsq}) by the $\pi$ form factor taken
from \cite{dubnicka}. The final result after the integration over the
allowed range of $M$ and translation into the branching fraction
is $(24.4\pm0.4)\%$. Our value is a little bigger than that of
K\"{u}hn and Santamaria \cite{santamar}, who used the same formula
but a different form factor, but still smaller than the experimental
value of $(25.24\pm0.16)\%$.

After consulting Fig.~\ref{taupipif} we can see that the differential
partial width of the decay $\tau^-\ra K^-K^0\nu_\tau$ can be obtained
from (\ref{taupipi}) by substituting $m_\pi\ra m_K$ and multiplying
by $(g_{\rho^-K^-K^0}/g_\rho)^2$. The latter quantity could only be
obtained from the analysis of the kaon electromagnetic form factor.
Here we  determine its product with $\Vud^2$ from the
experimental branching fraction. This product, denoted
as $X_{\ak K^-\rho^+}$,  will be used as an input parameter in
Sec.~\ref{psps}.

To get formulas for the $\tau^-$ decay rates into $\ksm$, $a_1^-$,
or $K_1^-$ mesons in narrow width approximation, we only need to
change the masses in (\ref{taurho})
and replace $\Vud$ by $w_{K^*}\Vus$, $w_{a_1}\Vud$, or
$w_{K_1}\Vus$, respectively.

Tsai \cite{tsai} assumed that the second Weinberg sum rule \cite{weinbsum}
is saturated by narrow-width $\rho$ and $a_1$ mesons and got the
prediction for the $\tau^-\ra a_1^-+\nu_\tau$ decay rate. In our
notation this situation would correspond to $w_{a_1}=1$. Here we treat
$w_{a_1}$ as a phenomenological parameter and determine its value
from the experimental branching fraction. The results is $w_{a_1}=
0.8044\pm0.0023$. The corresponding value of the parameter $Y_{a_1}$,
defined by Eq.~(\ref{ym}) is shown in Tab.~\ref{ymtab}.

In the case of $\tau^-\ra K_1^-\nu_\tau$ we proceed similarly and
obtain $w_{K_1}=0.84\pm0.21$.

The value $(w_{K^*}\Vus)^2=(4.20\pm0.09)\dek{-2}$ will be determined in
Sec.~\ref{pspslnu} from the experimental branching fraction of
$K^+\ra\pi^0e^+\nu_e$ and the full width of $\ksp$. The corresponding
$\tau^-\ra\ksm\nu_\tau$ branching fraction, which can be considered a
prediction of the MD approach, is shown in Tab.~\ref{tautabl}.

\subsection{Decays of the type \protect{$\tau^-\ra V+P+\nu_\tau$}}
\label{tauvp}

\begin{figure}
\begin{center}
\leavevmode
\setlength \epsfxsize{15cm}
\epsffile{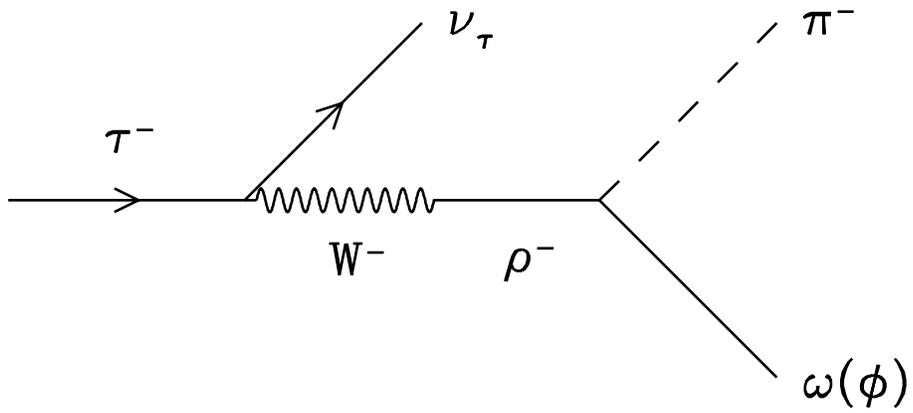}
\end{center}
\caption{Matrix element of the decay
$\tau^-\ra\omega (\phi)\ \pi^-\nu_\tau$.}
\label{tauompif}
\end{figure}

When dealing with decay $\tau^-\ra\omega\pi^-\nu_\tau$, see
Fig.~\ref{tauompif}, we have to exploit also the Lagrangian of
interaction among two vector fields and a pseudoscalar one
\be
\label{lagvvps}
{\cal L}_{V_1V_2P}=g_{V_1V_2P}\ \epsilon_{\mu\nu\rho\sigma}
\ \partial^\mu V_1^\nu\ \partial^\rho V_2^\sigma\ \varphi_P\ .
\ee
The differential decay width in masses of the $\omega\pi^-$ system
comes out as
\bea
\label{tauompi}
\frac{d\Gamma_{\tau^-\ra\omega\pi^-\nu_\tau}}{dM}&=&
\frac{(G_F\Vud)^2}{6(4\pi m_\tau)^3}
\left(\frac{g_{\rho^-\omega\pi^-}}{g_\rho}\right)^2\nl
&&\times\frac{1}{M^3}\left(m_\tau^2-M^2\right)^2
\left(m_\tau^2+2M^2\right)\ \lambda^{3/2}(M^2,m^2_\omega, m^2_\pi)
\left|F(M^2)\right|^2 \ ,
\eea
The $\rho^-\omega\pi^-$ coupling constant can be fixed using the VMD
and experimental branching fraction of the radiative decay
$\omega\ra\gamma\pi^0$, as discussed in Appendix~\ref{detccvvp}. It was
shown by
Decker \cite{decker} that a simple form factor like (\ref{fmsq}) did
not lead to a proper description of the $\omega\pi^-$ mass spectrum
\cite{argus87} and that a higher $\rho$ pole had to be included. He
used what was known at that time as $\rho(1600)$. Later it became
clear that the 1600 MeV region actually contains two $\rho$-like
resonances. Castro and L\'{o}pez \cite{castro} showed that
a better description of the same data \cite{argus87} is provided by
combining $\rho(770)$ with $\rho(1450)$ rather than with $\rho(1700)$.
Having fixed the admixture parameter, they obtained the branching
fraction that we included in our Tab.~\ref{tautabl} together with
the recently published experimental value \cite{buskulic}.

Formula (\ref{tauompi}) gives, after obvious modifications, also the
differential decay width of not yet observed decay mode
$\tau^-\ra\phi\pi^-\nu_\tau$. Here, the $\rho^-\phi\pi^-$ coupling
constant can be determined in a more direct way, namely, by exploring
the experimental branching fraction of $\phi\ra\rho\pi$ as shown in
Appendix~\ref{detccvvp}. Instead of performing our own analysis we
again quote the result of Castro and L\'{o}pez, who assumed that the
form factor is the same as in the $\omega\pi^-$ case.

\begin{figure}
\begin{center}
\leavevmode
\setlength \epsfxsize{15cm}
\epsffile{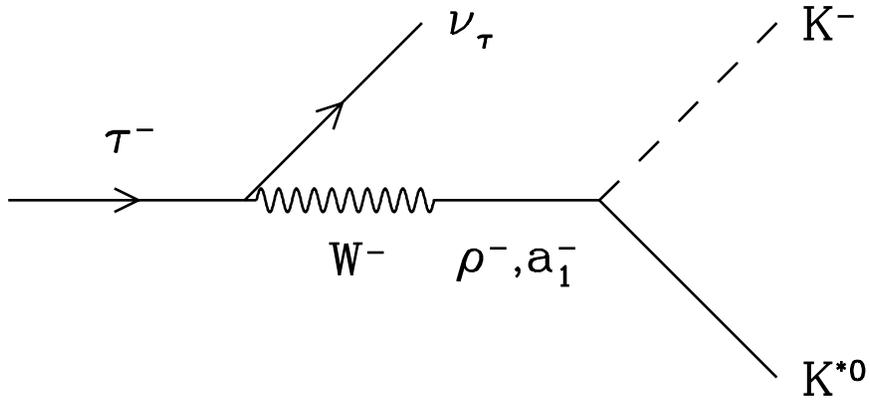}
\end{center}
\caption{Two diagrams that contribute to decay
$\tau^-\ra K^{*0} K^-\nu_\tau$, one with $\rho^-$, the other with
$a_1^-$ in the intermediate state.}
\label{taukstkf}
\end{figure}

The calculation of the branching fraction of the decay
$\tau^-\ra K^{*0}K^-\nu_\tau$ is complicated by the fact that two
Feynman diagrams, one with $\rho^-$,  the other with $a_1^-$ in the
intermediate state, see Fig.~\ref{taukstkf}, contribute to the transition
amplitude. The contribution of the former is proportional to the
$\rho^-K^{*0}K^-$ coupling constant, the value of which can be determined
by analyzing the $K^{*0}$ and $K^{*+}$ radiative decays by means of
the VMD in electromagnetic interactions. This analysis offers two
solutions for $(g_{\rho^-K^{*0}K^-}/g_\rho)^2$ that are compatible with
experimental data on the $K^*$ radiative decays, namely,
$(2.22\pm0.18)$~GeV$^{-2}$ and $(9.2\pm5.8)\dek{-2}$~GeV$^{-2}$,

If we forget for a moment about the axial current diagram
and calculate the branching fraction only from the diagram with
$\rho^-$ in the intermediate state, we find that it plays a
negligible role. Even for the larger solution shown above the resulting
branching fraction is very small, $(6.0\pm0.5)\dek{-5}$, far below
the experimental value of $(2.0\pm0.6)\dek{-3}$. It shows that the
dominant contribution is provided by the diagram with $a_1^-$ in
the intermediate state.\footnote{We must say that this conclusion
disagrees with that reached in Ref.~\cite{li}.} Unfortunately,
we do not have any possibility
to fix the $a_1^-K^{*0}K^-$ coupling constant. So instead of an honest
calculation let us make a crude estimate of what the experimental
information on the $\tau^-\ra K^{*0}K^-\nu_\tau$ and
$\tau^-\ra a_1^- \nu_\tau$ would imply if the former mode were a
subprocess of the latter. Dividing their branching fractions leads
to $B(a_1^-\ra K^{*0}K^-)\approx 1\%$. This value does not seem
to be excluded by the ``possibly seen" status of this mode in
\cite{pdg}.

\begin{figure}
\begin{center}
\leavevmode
\setlength \epsfxsize{15cm}
\epsffile{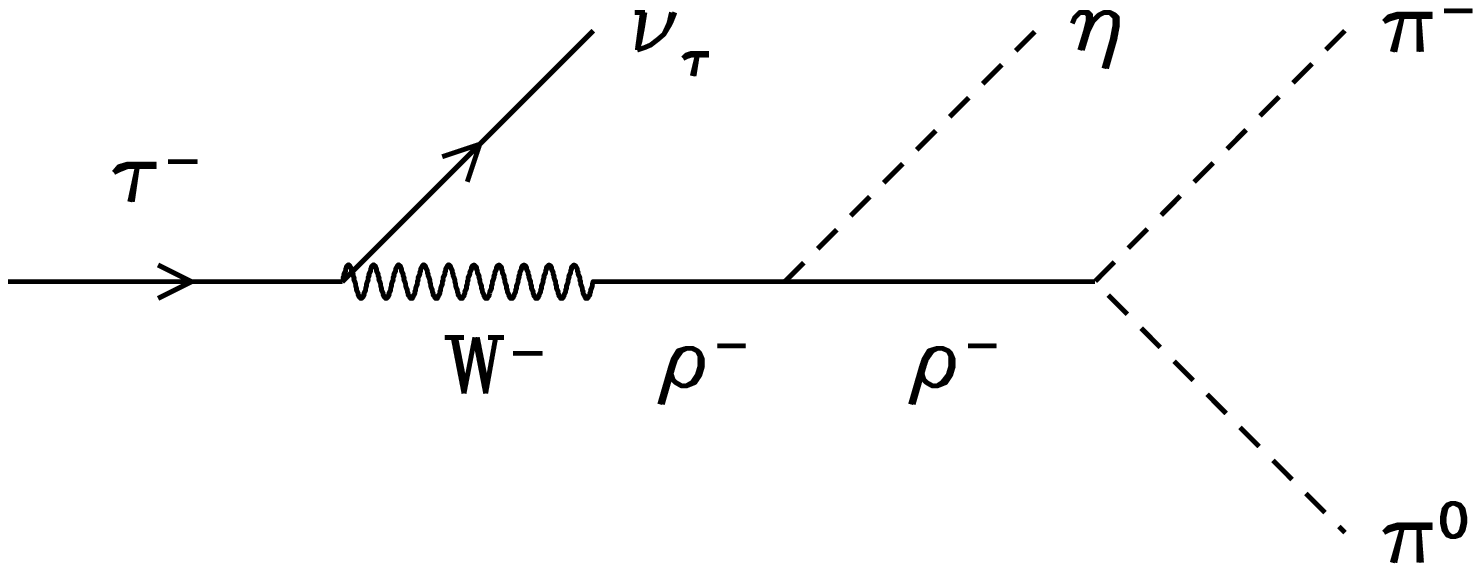}
\end{center}
\caption{Matrix element of the decay
$\tau^-\ra\eta\ \pi^-\pi^0\nu_\tau$.}
\label{tauetarf}
\end{figure}

Another example of the $\tau^-$ decay modes with one pseudoscalar and
one vector meson in final state is $\tau^-\ra\rho^-\eta\nu_\tau$.
This mode was considered a possible test of the Wess-Zumino term
\cite{wesszumi} for chiral anomalies \cite{chiralan}. The expected
branching fraction lay in the interval $(3.4,3.9)\dek{-4}$
\cite{kramer84}. In the MD approach we describe it by means of
the hadronic vertex connecting two $\rho$'s with the $\eta$, see
Fig.~\ref{tauetarf}. The proper interaction Lagrangian is again that
introduced in Eq.~\ref{lagvvps}. Because the narrow width approximation
is not as justified as well as it was in the case of the $\omega$ and
$\phi$ mesons, we complete the diagram with two pions originating from
the $\rho$ and evaluate the $\tau^-\ra\pi^-\pi^0\eta\nu_\tau$ branching
fraction. Everything greatly simplifies if we assume that the mass
difference between $\pi^-$ and $\pi^0$ can be neglected. The Feynman
diagram depicted in Fig.~\ref{tauetarf} then leads to
\bea
\label{taupipie}
\Gamma_{\tau^-\ra\pi^-\pi^0\eta\nu_\tau}&=&
\frac{(G_Fg_{\rho\rho\eta}\Vud)^2}{9(4\pi)^5m_\tau^3m_\rho^4}
\int_{2m_\pi}^{m_\tau-m_\eta}\ dM_2\ \left(M_2^2-4m_\pi^2\right)^{3/2}
\left|F(M^2_2)\right|^2\ \nl
&&\times\int_{M_2+m_\eta}^{m_\tau}
\frac{dM_1}{M_1^3}\left(m_\tau^2-M^2_1\right)^2
\left(m_\tau^2+2M^2_1\right)\lambda^{3/2}(M^2_1,M^2_2,m^2_\eta)
\left|F(M^2_1)\right|^2.
\eea
One yet unknown parameter is the coupling constant in the
$\rho\rho\eta$ vertex. As shown in Appendix~\ref{detccvvp}, the
branching fraction of $\rho^0\ra\eta\gamma$ can be utilized and the
value $(g_{\rho\rho\eta}/g_\rho)^2=(15.1\pm2.8)$~GeV$^{-2}$ is
obtained. Using the form factor \cite{dubnicka} and integrating
Eq.~(\ref{taupipie}) numerically we end up with the
branching fraction $(1.79\pm0.33)\dek{-3}$, which agrees perfectly
with the experimental value of $(1.71\pm0.28)\dek{-3}$.

\subsection{Decay \protect{$\tau^-\ra\pi^-\eta\nu_\tau$}}
\label{taupieta}
\begin{figure}
\begin{center}
\leavevmode
\setlength \epsfxsize{15cm}
\epsffile{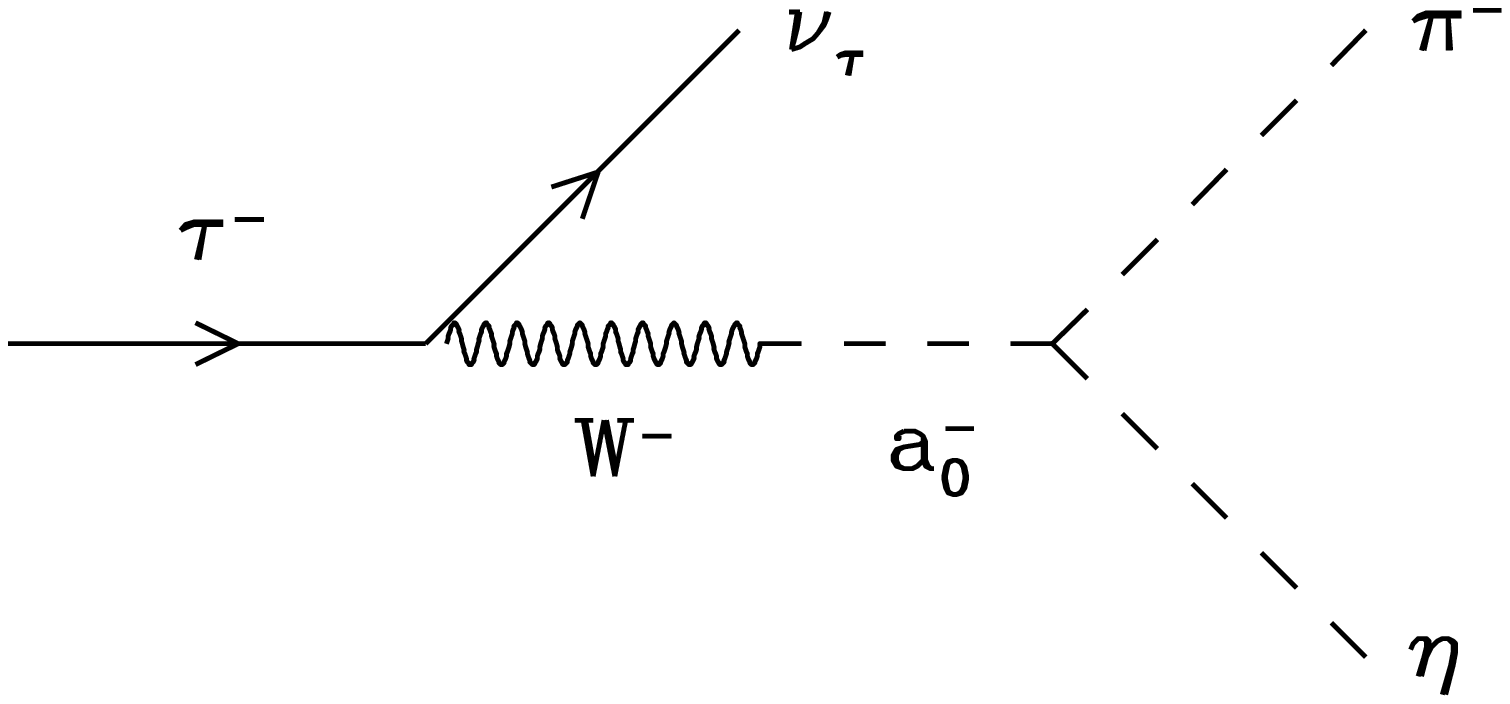}
\end{center}
\caption{Matrix element of the decay $\tau^-\ra\eta\ \pi^-\nu_\tau$.}
\label{tauetapi}
\end{figure}

The experimental upper limit for the branching fraction of the
$\tau^-\ra\pi^-\eta\nu_\tau$ mode ($1.4\dek{-4}$) indicates that this
mode is suppressed relative to $\tau^-\ra\rho^-\eta\nu_\tau$, which we
considered above, by at least one order of magnitude in spite of the
larger phase space available. This can easily be understood within the
MD approach. In fact, if the spin-parity conservation laws are strictly
enforced in conjunction with those based on isospin invariance, then
there is no pseudoscalar, vector, or axial-vector meson that can couple
to the $\pi^-\eta$ system. The only possibility to realize the
transition from $W^-$ to this system is via the $a_0^-$ scalar meson,
see Fig.~\ref{tauetapi}. It gives us a chance to gain some information
about the strength of the $a_0^-W^-$ interaction, as shown in the
following.

The dominant decay mode of the $a_0^-$ meson is $a_0^-\ra\pi^-\eta$.
It allows us to  replace the decay $\tau^-\ra\pi^-\eta\nu_\tau$ in our
considerations by a simpler one, namely,
$\tau^-\ra a_0^-\nu_\tau$.\footnote{This decay was proposed as a clear
test for the existence of second-class vector current by Leroy and
Pestieau \cite{leroy} soon after the discovery of the $\tau$ lepton
\cite{perl}.} The coupling of scalar mesons to the gauge bosons is
similar to that of pseudoscalar mesons, as shown by a comparison of
(\ref{pslag}) and (\ref{slag}). We can therefore use Eq.~(\ref{taups})
to find an upper bound on the scalar-meson decay constant of the
$a_0^-$ meson. The result is $f_{a_0^-}<7$~MeV. For comparison, the
pseudoscalar decay constant of the $\pi^-$ meson is about 131 MeV.

\section{PROCESSES CONTAINING THE $P-P-V^+$ VERTEX WITH $V^+$
COUPLED TO $W^+$}
\label{psps}
In this section we will consider charged weak current decays of
pseudoscalar mesons ($P_1$) into a pseudoscalar meson ($P_2$) and an
additional system, which may be an $\ell\nu$ pair, another pseudoscalar
meson, a vector meson, or an axial-vector meson. According  to the
MD hypothesis the processes of this kind proceed by coupling the
pseudoscalar meson pair to a charged vector meson ($V$), which in turn
couples to a charged gauge boson. The latter finally converts into one
of the systems mentioned above. The $P_1\ra P_2+V$ transition is
governed by the Lagrangian (\ref{lagvpsps}). It is useful to introduce
the quantity
\be
\label{x}
X_{P_1P_2V}=Y_V\left(\frac{g_{VP_1P_2}}{g_\rho}\right)^2\ ,
\ee
which will enter all our formulas for decay rates in this section.
Parameter $Y_V$ is defined by Eq.~(\ref{ym}), $g_{VP_1P_2}$ is the
coupling constant in the Lagrangian (\ref{lagvpsps}).

Our general strategy will be to determine the quantities (\ref{x}) from
some of the experimentally known branching fractions of semileptonic
decays and then use them for making predictions for other decay modes.
A notable exception is  the decay $\pi^+\ra\pi^0 e^+ \nu_e$, which
proceeds via the $\rho^+$ meson. Here, the quantity under consideration
is simply given by the $ud$ element of the CKM matrix,
$X_{\pi^+\pi^0\rho^+}=\Vud^2$, and is thus well known.

\subsection{Decays of the type $P_1 \ra P_2 +\ell^+\nu_\ell$}
\label{pspslnu}
\begin{figure}
\begin{center}
\leavevmode
\setlength \epsfxsize{15cm}
\epsffile{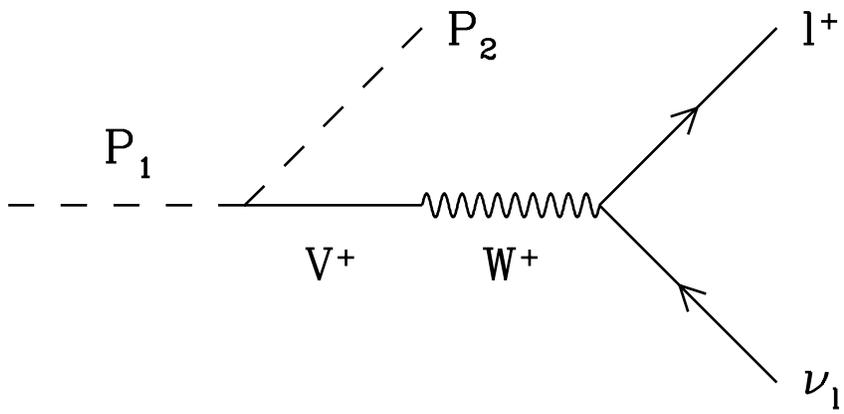}
\end{center}
\caption{Generic Feynman diagram of $P_1\ra P_2 \ell^+\nu_\ell$
decays.}
\label{pspslnuf}
\end{figure}

The generic Feynman diagram for the weak decay of a pseudoscalar meson
$P_1$ into another pseudoscalar meson $P_2$ and an $\ell^+\nu_\ell$
pair is shown in Fig.~\ref{pspslnuf}. The corresponding matrix element
can easily be written on the basis of the lepton part of the standard
model Lagrangian and Eqs. (\ref{ccvavd}) and (\ref{lagvpsps}).
\be
\label{mpspslnu}
{\cal M}=G_Fw_{V}V_V\frac{g_{VP_1P_2}}{g_\rho}\frac{m_V^2}
{m_V^2-t}\left[(p_1+p_2)^\mu-\frac{m_{P_1}^2-m_{P_2}^2}{m_V^2}
(p_1-p_2)^\mu \right]{\overline l}\gamma_\mu(1-\gamma_5)\nu\ ,
\ee
where $p_1$ ($p_2$) is the four-momentum of the incoming (outgoing)
meson and $t=(p_1-p_2)^2$ is the square of the four-momentum
transfer from $P_1$ to $P_2$. Obviously, $t$ is also equal to
the mass squared of the $\ell^+\nu_\ell$ system. The evaluation based
on Eqs.~(\ref{3body}) and (\ref{mpspslnu}) gives, after the integration
over the lepton momentum direction in the $\ell^+\nu_\ell$ rest frame,
the following formula for the differential partial width:
\be
\label{dpspslnu}
\frac{d\Gamma_{P_1\ra P_2\ell^+\nu_\ell}}{dt}=\frac{G_F^2X_{P_1P_2V}}
{3(4\pi m_{P_1})^3}\ \frac{t-m_\ell^2}{t^3}\ \lambda^{1/2}
(m_{P_1}^2,m_{P_2}^2,t)
\left[\varphi_1(t)-
\varphi_2(t)\right]\left(\frac{m_V^2}{m_V^2-t}\right)^2\ ,
\ee
where
\bea
\label{phi1}
\varphi_1(t)&=&2t^4-(4x+4y+z)t^3+\left[2(x-y)^2+z(2x+2y-z)
\right]t^2\nl
&&+2z\left[(x-y)^2+z(x+y)\right]t  - 4z^2(x-y)^2\ , \nl
\varphi_2(t)&=&\frac{3tz}{r^2}(2r-t)(t-z)(x-y)^2 .
\eea
We used the notation $r=m_V^2$, $x=m_{P_1}^2$, $y=m_{P_2}^2$, and
$z=m_\ell^2$.
Integrating Eq. (\ref{dpspslnu}) in case of the $\pi^+\ra\pi^0e^+\nu_e$
decay (usually referred to as $\pi_{e3}$),  we get a branching fraction
of $(1.0041\pm0.0021)\times10^{-8}$. The error comes from the mean
$\pi^\pm$ lifetime we used to convert the decay rate into the branching
ratio and from the $V_{ud}$ element of the CKM matrix. The agreement
between our result and the experimental value of
$(1.025\pm0.034)\times10^{-8}$ is perfect.

As it has already been mentioned, the $\pi_{e3}$ decay is exceptional
because the coupling constant in the hadronic vertex just cancels
with the $g_\rho$ coming from the MD Lagrangian (\ref{ccvavd}) and the
$w_{\rho^+}$ parameter is exactly one, as it follows from the
normalization of the pion electromagnetic form factor. In other
semileptonic decays the formulas for branching fractions contain badly
known or unknown parameters $X_{P_1P_2V}$ defined in Eq.~(\ref{x}).
Some experimental branching fractions have been explored to determine
those parameters, with resulting values shown in Table~\ref{xtab}.
Others, shown in Tab.~\ref{pspslnut}, provide the check of the
soundness of the MD results.

For example, the branching fraction of the decay $K^+\ra\pi^0 e^+\nu_e$
is used to fix the value of $X_{K^+\pi^0\ksp}$ at
$(1.206\pm0.015)\dek{-2}$; those of $K^+\rightarrow\pi^0 \mu^+\nu_\mu$,
$K^0_L\ra\pi^\pm e^\mp\bar\nu_e(\nu_e)$, and
$K^0_L\ra\pi^\pm\mu^\mp\bar\nu_\mu(\nu_\mu)$ then come as predictions of
the MD. The result for electron mode of $K_L^0$ is somewhat higher than
the experimental value. It may signal the presence of isospin symmetry
violating effects \cite{leutroos,gassleut}.
After taking the experimental value of $\Vus$ and
determining the coupling constant ratio $g^2_{\ksp K^+\pi^0}
/g_\rho^2=0.2872\pm0.0051$ from the $\rho$ and $K^{*+}$ decay widths
we isolate $w_{\ksp}=0.929\pm0.013$. The deviation of the latter
from unity is  what one would expect for an $SU_f(3)$ breaking effect.

For other decay modes such a detailed analysis cannot be performed
because the hadronic coupling constants of vector meson resonances are
either inaccessible for fundamental reasons (e.g., hadronic decay is not
kinematically allowed) or because the decay widths are poorly known.
We are thus left with the $X_{P_1P_2V}$ values shown in Tab. \ref{xtab},
without the possibility to extract the $w_V$ parameters. But it does not
hamper our ability to predict the branching fractions of related
processes $P_1\ra P_2+P_3$ and $P_1\ra P_2+V(A)$.

A very interesting situation is in the semileptonic decays of
$B$-mesons. Frequent decay modes $B^0\ra D^-\ell^+\nu_\ell$,
$B^+\ra \ad\ell^+\nu_\ell$, and $B_s^0\ra D_s^-\ell^+\nu_\ell$ cannot
be explained within the MD framework without assuming the existence of
a vector meson with both charm and beauty, $\bcsp$.  But such a meson
has not yet been discovered experimentally. In order to proceed further
we simply assume that it does exist and choose its mass at
6.34~GeV/$c^2$, as determined by Godfrey and Isgur \cite{potent1} in
a relativized quark model with chromodynamics. This value agrees with
results of other potential models \cite{potent2}. We will return to
this question, which is of vital importance for the MD hypothesis,
in Sec.~\ref{comments}.

In Tab.~\ref{pspslnut} we also show the predictions for semileptonic
decay modes with the $\tau$ lepton.\footnote{The issue of applicability
the MD approach to decays of heavy mesons will be addressed in
Sec.~\ref{heavy}.} It is natural to ask to which extent the branching
ratio of the $\tau$ and light lepton modes can discriminate
among various models. Let us mention that for the $B^+\ra\pi^-$
semileptonic decays  our ratio is 0.52, while the recent estimate
by Khodjamirian and R\"{u}ckl \cite{khodruck} is 0.7--0.8.

\begin{figure}
\begin{center}
\leavevmode
\setlength \epsfxsize{15cm}
\epsffile{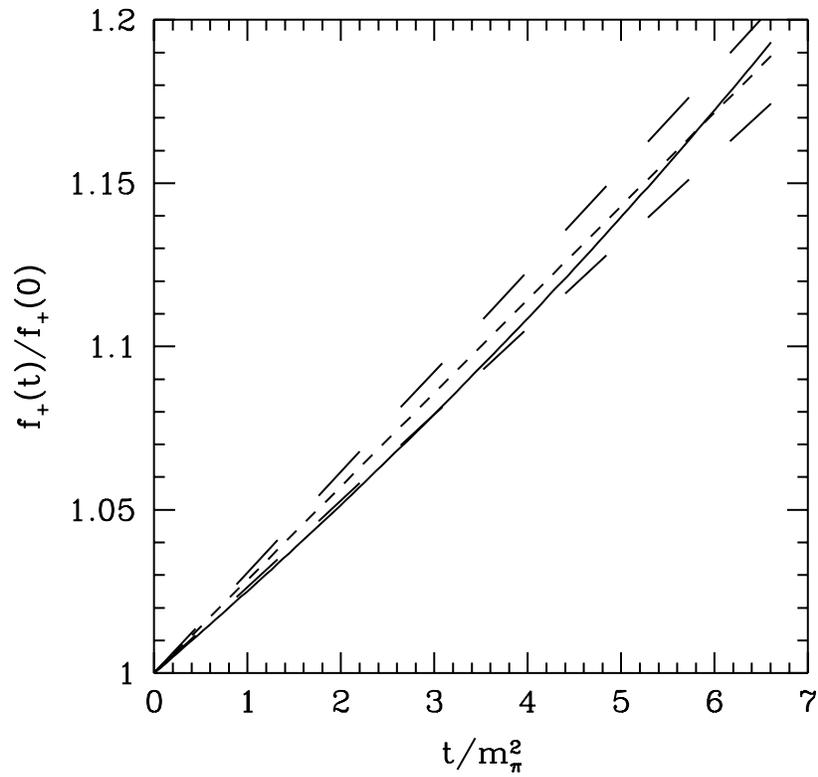}
\end{center}
\caption{$K^+_{e3}$ form factor $f_+(t)$: Meson dominance (solid),
linear parametrization used by experimentalists to fit data with limits
coming from the experimental error of the slope parameter (dashed).}
\label{kformfac}
\end{figure}

The branching fractions are not the only outcome of the MD approach that
can be compared with experimental data. Any experimentally observable
quantity can be calculated. Let us consider as an example the form
factors $f_+(t)$ and $f_-(t)$ of the $K_{e3}^+$ decay, which are
defined by
\be
\label{ffdef}
{\cal M}=\frac{G_FV_{us}}{\sqrt{2}}
\left[f_+(t)(p_1+p_2)^\mu+f_-(t)(p_1-p_2)^\mu\right]
\bar\ell\gamma_\mu(1-\gamma_5)\nu \ .
\ee
Differential decay rate for arbitrary form factors is shown in
Appendix~\ref{genff}. If we now compare Eq.~\ref{mpspslnu} with the
definition (\ref{ffdef}) of $f_+(t)$, we see that in the MD approach the
$t$-dependence of the latter is given by
\be
\label{fplus}
f_+(t)=f_+(0)\frac{m^2_{\ksp}}{m^2_{\ksp}-t}\ .
\ee
The comparison with the linear parametrization used by experimentalists
is shown in Fig.~\ref{kformfac}. A faster than linear rise of the MD
form factor may explain why the experimentally determined slope
parameters in the $K_{\mu3}$ decays are higher than those in $K_{e3}$.

Furthermore, from (\ref{mpspslnu})
we are getting the form factor ratio
\be
\label{fminus}
\frac{f_-(t)}{f_+(t)}=-\frac{m_{K^+}^2-m^2_{\pi^0}}{m^2_{\ksp}}=
-0.28367\pm0.00016\ ,
\ee
which should be compared to the experimental value of $-0.35\pm0.15$.
Equations (\ref{fplus}) and (\ref{fminus}) were first derived by Dennery
and Primakoff \cite{dennery}, who saturated the dispersion relations
for the form factors by a $K^*$ pole.\footnote{In fact, they considered
two vector resonances. One of them, $K^*(730)$, was abandoned later on.}

\subsection{Decays of the type $P_1 \ra P_2 + P_3$}
\label{pspsps}

\begin{figure}
\begin{center}
\leavevmode
\setlength \epsfxsize{15cm}
\epsffile{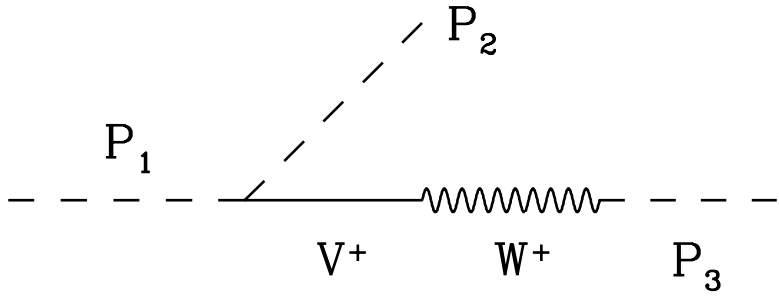}
\end{center}
\caption{Generic Feynman diagram of $P_1\ra P_2+P_3$ decays.}
\label{pspspsf}
\end{figure}

The generic Feynman diagram of the processes we are going to consider
now is shown in Fig.~\ref{pspspsf}. The parent pseudoscalar meson
$P_1$ undergoes the strong interaction conversion into $P_2$, one of
the outgoing pseudoscalar mesons, and a charged (virtual) vector meson
$V$. The latter couples according to (\ref{ccvavd}) to the $W$ boson,
which in turn converts to the second outgoing pseudoscalar meson $P_3$.
To simplify the discussion we will consider only positively charged
$V$'s, which means neutral or positively charged $P_1$'s and,
correspondingly, negatively charged or neutral $P_2$'s. This convention
clearly shows which of the two final-state pseudoscalar mesons is
coupled to the gauge boson. Of course, we automatically handle
also the charge conjugate modes.

The mechanism considered here does not operate in all $P_1\ra P_2+P_3$
transitions. It cannot explain any of the decays into two neutral
mesons. Also some charged modes cannot run in this way. Let us take
as an example $D_s^+\ra\ak K^+$. There does not exist any vector meson
that would appear together with $\ak$ as a result of the strong
conversion of $D_s^+$. And what accompanies $K^+$ in such a conversion
is $D^{*0}$, which cannot couple to any of the weak gauge bosons.

The partial decay width comes out from the Feynman diagram in
Fig.~\ref{pspspsf} as
\be
\label{pspspsga}
\Gamma_{P_1\ra P_2+P_3}=\frac{G_F^2}
{16\pi m_1^3}\ X_{P_1P_2V}Z_{P_3}
\left(m_{P_1}^2-m_{P_2}^2\right)^2\ \lambda^{1/2}(m_{P_1}^2,m_{P_2}^2,
m_{P_3}^2)\ .
\ee
Parameters $X_{P_1P_2V}$, defined by Eq.~(\ref{x}), have already
been assigned numerical values using experimental information on
some semileptonic decay modes, as shown in Table~\ref{xtab}. Similarly,
the parameters $Z_{P_3}$ are defined by Eq.~(\ref{z}).
Their values were determined from the leptonic branching fractions
of pseudoscalar mesons and are shown in Table~\ref{fvtab}.

The results obtained from (\ref{pspspsga}) for various input
and output mesons were converted to branching fractions by means of
experimental lifetimes. They can be divided into three groups. In
Tab.~\ref{pspspstg} we present calculated branching fractions that
agree with experimental data. Their less lucky companions are listed
in Tab.~\ref{pspspstb}. We defer the discussion about possible meaning
of discrepancies between our results and empirical values to
Sec.~\ref{comments}. The last group, shown in Tab.~\ref{pspspstn},
comprises the branching fractions
that have not been measured yet. When the experimental information
becomes more complete some of the modes listed there may fall into the
first category, some into the second one.

\subsection{Decays of the type $P_1 \ra P_2 + V (A)$}
\label{pspsv}

\begin{figure}
\begin{center}
\leavevmode
\setlength \epsfxsize{15cm}
\epsffile{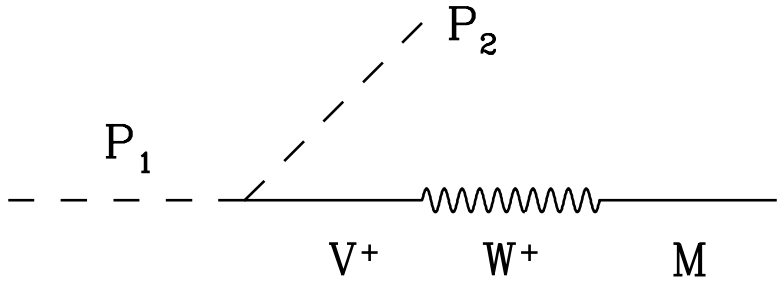}
\end{center}
\caption{Generic Feynman diagram of $P_1\ra P_2+M$ decays. $M$ stands
for the outgoing vector or axial-vector meson.}
\label{pspsvf}
\end{figure}

Keeping in mind our convention about charges of the parent pseudoscalar
mesons, the flavor-changing decays we are going to analyze now can
proceed in the lowest order of the MD approach only through the diagram
depicted in Fig.~\ref{pspsvf}. Because the vector ($V$) and axial-vector
($A$) mesons couple to the charged gauge bosons in the same way, we can
study the two modes, one with an outgoing vector meson, the other with
an outgoing axial-vector meson, simultaneously. We will label either of
those two mesons as $M$, freeing the index $V$ for the intermediate
vector meson that connects the hadronic vertex with $W^+$. The partial
decay width summed over the spin projections of $M$ reads
\be
\label{pspsvga}
\Gamma_{P_1\ra P_2+M}=\frac{G_F^2m_M^2 X_{P_1P_2V}\ Y_M}
{8\pi g_\rho^2m_{P_1}^3}\left(\frac{m_V^2}
{m_V^2-m_M^2}\right)^2\ \lambda^{3/2}(m_{P_1}^2,m_{P_2}^2,m_M^2)\ .
\ee
The branching fractions calculated from (\ref{pspsvga}) and mean life
times of the parent mesons ($D^+$, $D^0$, $D_s^+$, $B^+$, and $B^0$)
are shown in Tables \ref{pspsvtg} and \ref{pspsvtb} together with
experimental values.
The former table lists the decay modes for which the MD results do not
contradict the experiment. As most of the empirical values are given
only as upper bounds at present, some modes may move in future to
Table~\ref{pspsvtb}, which contain the MD results that disagree with
the data.

\subsection{Meson dominance and decays of heavy mesons}
\label{heavy}

The decays rates of heavy mesons containing a heavy quark and a light
antiquark, or vice versa, are usually calculated using the heavy
quark effective theory \cite{HQET}. The careful reader has probably
noticed when inspecting Tables \ref{pspslnut}-\ref{pspsvtb} that we
used the MD formulas to calculate also the decay fractions of the heavy
mesons $D$, $D_s$, $B$, and $B_s$. It seems to go against the spirit
of the MD, as it has been declared in Sec.~\ref{deriv}, because the
energies in the parent rest frame of the outgoing particles are large.
But, as we will argue, the use of MD for calculating the branching
ratios of different light meson modes is well justified.

In fact, what matters is the virtuality (four-momentum squared) flowing
through the junction where a meson and the weak gauge boson meet. Let's
speak, for definiteness, about the decay of $B^0$ into $D^-$ and a
light meson $M^+$ (where now $M$ can be $P$, $V$, or $A$), which is
supposedly dominated by the $\bcsp$. See Figs. \ref{pspspsf} and
\ref{pspsvf}. Because the meson $M^+$ is on the mass shell, the
virtuality in the $W^+M^+$ junction is equal to the mass of $M$ squared,
i.e., it is same as in the definitions of the weak effective Lagrangians
in Sec.~\ref{deriv}. The story at the opposite end of the $W^+$ line,
in the $\bcsp W^+$ junction, is different.  Here, the virtuality is far
from the $\bcsp$ mass squared, and the coupling parameter may be
different from the $w_{\bcsp}$, which is defined for the on-shell
$\bcsp$. Similar changes may occur in the strong $B^0 D^-\bcsp$ vertex,
where again
the dependence on the virtualities of participating mesons cannot be
ruled out. The absolute predictions of the decay rates may thus be
unreliable. But we cannot make the absolute predictions anyhow because
we know neither $w_{\bcsp}$ nor $g_{B^0 D^-\bcsp}$, which combine into
$X_{B^0 D^-\bcsp}$, see Eqs.~(\ref{ym}) and (\ref{x}). We can only
calculate the ratios of decay rates for different mesons $M^+$.
When these mesons are light, the differences among the virtualities
in the $W^+\bcsp$ junction will be small in comparison with the mass
of the $\bcsp$ meson. The virtuality modified coefficients
$X_{B^0 D^-\bcsp}$ will have approximately the same value and will simply
cancel out when a ratio of the decay rates is calculated. The same happens
when a light meson mode is compared to the $\ell^+\nu_\ell$ one.

To push things to the edge we calculated the branching ratios of various
decay modes of the $B_c$ meson, containing both heavy valence quark and
antiquark. This meson has not been discovered yet, but reliable
calculations of its mass and other properties exist in the literature,
see, e.g., \cite{potent1,potent2}. Also the prospects of its impending
discovery are bright, see \cite{bcstar} and references therein.
Following Godfrey and Isgur \cite{potent1} we used in our calculations
the $B_c$ mass of 6.27~GeV/$c^2$.  The results of the MD approach are
compared to the predictions of some existing more fundamental models in
Table~\ref{comp}.

\subsection{Meson-electron induced binary reactions}
\label{mesonel}
\begin{figure}
\begin{center}
\leavevmode
\setlength \epsfxsize{15cm}
\epsffile{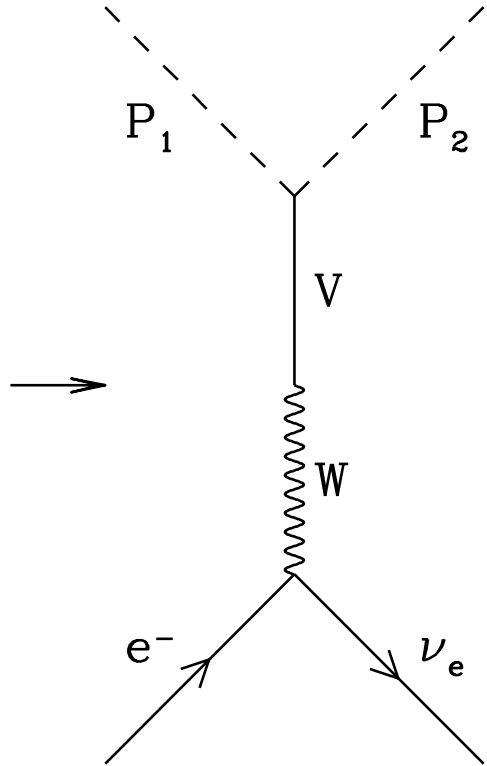}
\end{center}
\caption{Matrix element of the reaction $P_1^++e^-\ra P_2^- +\nu_e$.}
\label{psepsnuf}
\end{figure}

\begin{figure}
\begin{center}
\leavevmode
\setlength \epsfxsize{15cm}
\epsffile{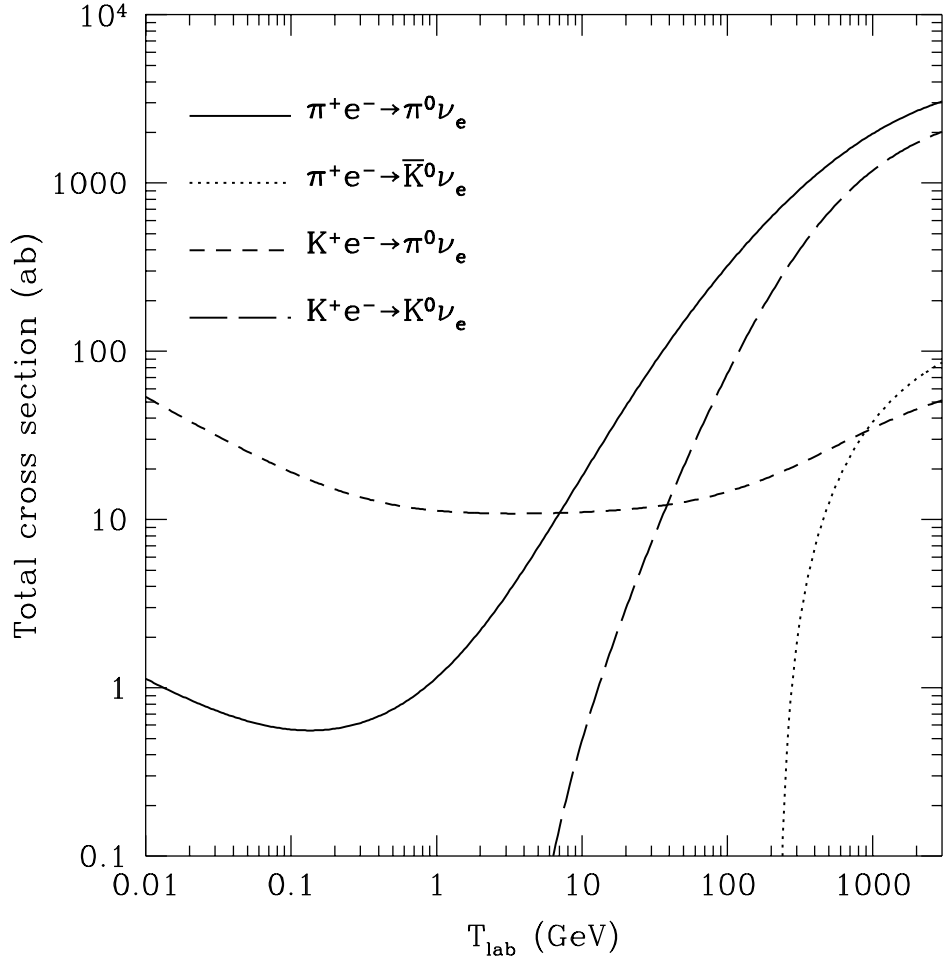}
\end{center}
\caption{Total cross section in attobarns (1 ab = $10^{-42}$~cm$^2$)
of the reactions of positive pions and kaons with
target electrons as a function of the laboratory kinetic energy.}
\label{psepsnug}
\end{figure}

Let us consider the following weak-interaction binary reactions of
projectile mesons incident on  target electrons:
$\pi^+e^-\ra\pi^0\nu_e$, $K^+e^-\ra\pi^0\nu_e$,
$\pi^+e^-\ra\ak\nu_e$, and $K^+e^-\ra K^0\nu_e$. First two of them
are exoenergetic, whereas the laboratory kinetic energy thresholds for
the remaining two are 223.1 GeV and 3.381 GeV, respectively. As a
consequence of the special kinematics (electron as a target), the
reactions remain in the low center-of-mass energy range
($s, |t|<1$~GeV$^2$) even for the highest meson beam energies
available. With view toward successful description of the semileptonic
decays of pion and kaons, we believe that the MD approach is suitable
also for calculating the cross sections of the low-energy reactions that
are related to those decays by crossing symmetry. The corresponding
Feynman diagram is depicted in Fig.~\ref{psepsnuf}. To make the
differential cross section formula concise, we introduced $x=m_{P_1}^2$,
$y=m_e^2$, $z=m_{P_2}^2$, and $r=m_V^2$. The formula then reads
\be
\label{psepsnu}
\frac{d\sigma_{P_1e\ra P_2\nu}}{dt}=\frac{G_F^2X_{P_1P_2V}
(\hbar c)^2}{8\pi\lambda(s,x,y)}\left(\frac{r}{r-t}\right)^2
\left[\phi_1(s,t)+\frac{x-z}
{r^2}\phi_2(s,t)\right]\
\ee
with
\bea
\phi_1(s,t)&=&4s(s-x-y-z)+4xz+y^2+(4s-y)t\nl
\phi_2(s,t)&=&2ry(2s+t-y-2z)+y(x-z)(y-t) \ .\nonumber
\eea
Total cross section as a function of the kinetic energy of incident
meson for the four meson-electron reactions mentioned above were
obtained by numerical integration and are shown in Fig.~\ref{psepsnug}.

\subsection{Antineutrino-electron induced mesonic reactions}
\label{antinuel}

\begin{figure}
\begin{center}
\leavevmode
\setlength \epsfxsize{15cm}
\epsffile{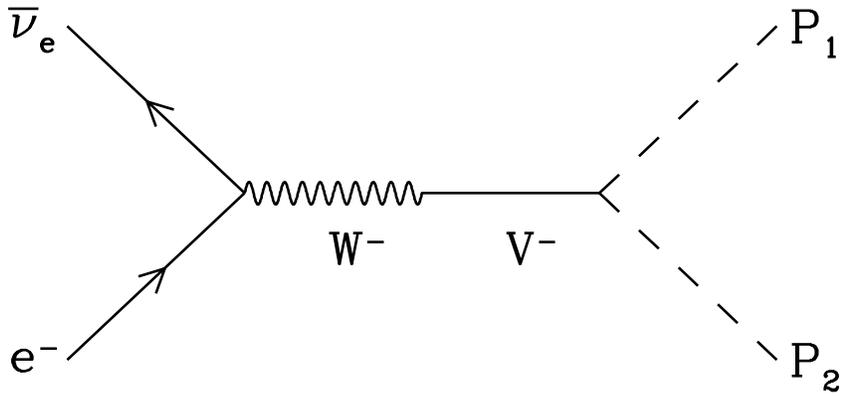}
\end{center}
\caption{Matrix element of the reaction $\bar\nu_e+e^-\ra P_1+P_2$.}
\label{nuepspsf}
\end{figure}

\begin{figure}
\begin{center}
\leavevmode
\setlength \epsfxsize{15cm}
\epsffile{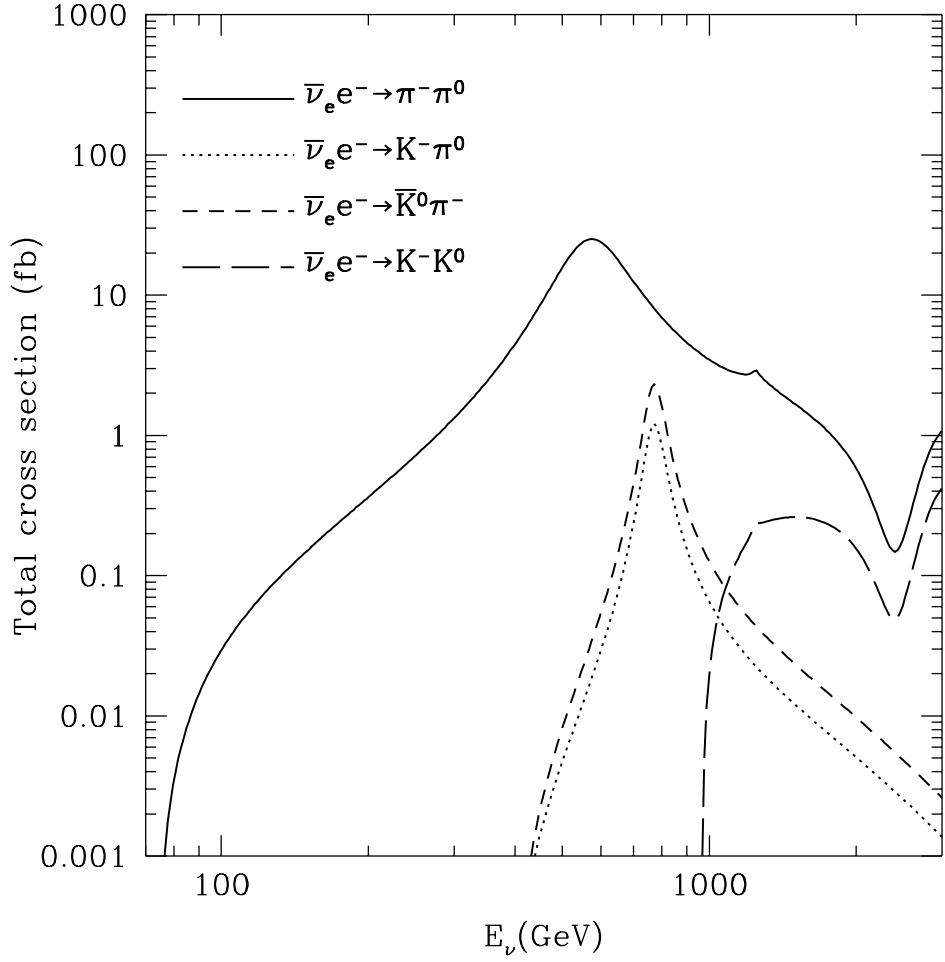}
\end{center}
\caption{Total cross section in femtobarns (1 fb = $10^{-39}$~cm$^2$)
of two-meson production in reactions of the electron antineutrino with
target electrons as a function of antineutrino energy.}
\label{nuepspsg}
\end{figure}

The electron antineutrino energies required for meson-pair production
off target electrons are very high. We show four most favorable final
states with the threshold antineutrino energies in parentheses:
$\pi^-\pi^0$ (73.8 GeV), $K^-\pi^0$ (387 GeV), $\ak\pi^-$ (398 GeV),
and $K^-K^0$ (962 GeV). Also the fact that the electron antineutrinos
are less copious than the muon ones by a factor of $10^{-4}$ in the high
energy antineutrino beams, produced by the $\pi^-_{\ell2}$ decays, makes
the experimental observation of this kind of reaction tricky. On the
other hand, as the cms energy of the two-meson  system remains small,
the transverse momenta of outgoing mesons will also be small. The
reaction products will thus be concentrated in a very narrow cone in the
laboratory system polar angle with relatively small energy spread. This,
and also the negative total charge, may help to identify this kind of
reactions. The evaluation of the total cross section that corresponds to
the Feynman diagram in Fig.~\ref{nuepspsf} gives the result
\bea
\label{nuepsps}
\sigma_{\bar\nu_e+e^-\ra P_1+P_2}&=&\frac{G_F^2(\hbar c)^2X_{P_1P_2V}}
{24\pi r^2s^3(s-y)^2}\left|F(s)\right|^2
\lambda^{1/2}(s,x,z) \nl
&&\times\left[2r^2s^3\lambda(s,x,z)+2r^2y^2(2y-3s)(x-z)^2+r^2sy
(3s^2-y^2)\right.\nl
&&\times\left.(2x+2z-s)+3sy(x-z)^2(s-y)^2(s-2r)\right]\ ,
\eea
where $x=m_{P_1}^2$, $y=m_e^2$, $z=m_{P_2}^2$, $r=m_V^2$, and
\be
\label{fssq}
\left|F(s)\right|^2=\frac{m_V^4}{\left(s-m_V^2\right)^2
+m_V^2\Gamma_V^2}\ .
\ee
For channels with the $\rho$ resonance in the $s$-channel we replaced
function (\ref{fssq}) by the form factor taken from
Ref.~\cite{dubnicka}, in which the experimental data on
$\die\ra\pi^+\pi^-$ were fit with a formula exhibiting the
correct analytic behaviour. In this way we have accounted for a possible
contribution from higher $\rho$ recurrences. For reactions with a
$\pi K$ system in the final state, which go through the $K^{*}$
resonance in the $s$-channel, we do not have such a possibility. The
single-pole formula (\ref{fssq}) with energy dependent $K^*$ width was
used.  The dependence of the total cross section on the incident
antineutrino energy for all four final states is shown in
Fig.~\ref{nuepspsg}.

\section{MESON DOMINANCE AND NEUTRAL FLAVOR CHANGING DECAY MODES}
\label{neutral}

The processes we are going to deal with now are usually classified
\cite{pdg} as flavor changing ($\Delta S=1$,  $\Delta C=1$) weak neutral
current decay modes. This label is a little misleading for some of them,
e.g., $K^+\ra\pi^+\dil$. In the calculations based on the standard model
the latter is described in terms of diagrams that almost all contain
\cite{vain76} the charged gauge boson $W^\pm$, i.e., the charged weak
current. Also in the MD approach we will calculate branching fractions
of this and similar decay modes using the diagrams where charged mesons
are attached to the charged weak gauge bosons. Only in a part of this
class of processes ($K^+\ra\pi^+\bar\nu\nu$, for example) the genuine
weak neutral current operates in conjunction with the charged one, which
can only change the flavor.

We start with considering the decay mode $K^+\ra\pi^+\dil$, which
was investigated theoretically already before its discovery in 1975
\cite{bloch}. References to this early period can be found in
\cite{vain76}. Later works include \cite{ecker87,berg,fajfer96} and
references therein. Present theoretical understanding of this decay
in the framework of chiral perturbation theory has recently been
summarized in \cite{pich96,ecker96}. The theoretical prediction
based on \cite{ecker87} contains one unknown parameter. When extracting
it from the experimental branching fraction of the dilectron mode, a
two-fold ambiguity remained. It was resolved by choosing the solution
that fit the $\die$ mass spectrum \cite{alliegro} better. Then the
prediction for the $\dim$ mode can be made.

\begin{figure}
\begin{center}
\leavevmode
\setlength \epsfxsize{15cm}
\epsffile{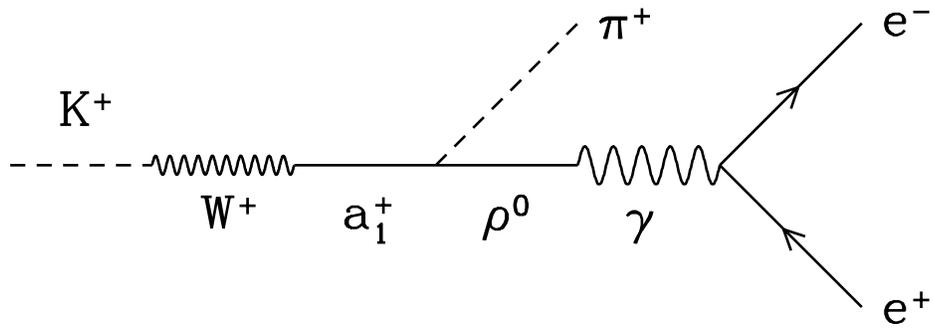}
\end{center}
\caption{Matrix element of the decay $K^+\ra\pi^+e^+e^-$ in
the meson dominance approach.}
\label{kpieef}
\end{figure}

In the MD approach we will describe the decay $K^+\ra\pi^+\dil$
by the diagram sketched in Fig.~\ref{kpieef}, forgetting for a while
about other possible diagrams. In order to evaluate the corresponding
decay rate we assume that the interaction among  the $a_1$, $\rho$,
and $\pi$ mesons is governed by the Lagrangian density
\begin{equation}
\label{a1lag}
{\cal L}_{a_1\rho\pi}=i\ g_{a_1\rho\pi}
\sum_{i,j,k}C_{i;j,k}\  V^\dagger_{j\alpha\beta}
\ \partial^\alpha\varphi^\dagger_k\ A_i^\beta
\end{equation}
with
\begin{equation}
\label{vst}
V_{j\alpha\beta}=
\partial_\alpha V_{j\beta}-\partial_\beta V_{j\alpha}
\end{equation}
being the vector field strength tensor. Symbols $A_i$, $V_j$, and
$\varphi_k$ denote the field operators for $a_1$, $\rho$, and $\pi$
mesons, respectively. Italic indices label various charge states of
a particular meson and $C_{i;j,k}$ is the SU(2) Clebsch-Gordan
coefficient. The $a_1\ra\rho\pi$ decay width comes from (\ref{a1lag})
as
\be
\label{a1width}
\Gamma_{a_1\rightarrow \rho+\pi}=\frac{g^2_{a_1\rho\pi}}
{96\pi m_{a_1}^3}\ \lambda^{1/2}(x,y,z)
\bigg[(x-y-z)^2+\frac{y}{2x}(x-y+z)^2\bigg] \ ,
\ee
where $x=m_{a_1}^2$, $y=m_\rho^2$, and $z=m_\pi^2$.
Substituting the experimental value of $\Gamma_{a_1}\approx 400$~MeV
into Eq.~(\ref{a1width}) we get
$g^2_{a_1\rho\pi}\approx 260$~GeV$^{-2}$.

A straightforward evaluation of the diagram depicted in Fig.~\ref{kpieef}
leads to the following formula for the differential width in the $\dil$
mass $M$
\bea
\label{kpiee}
\frac{d\Gamma_{K^+\ra\pi^+\dil}}{dM}&=&
\frac{(G_Fg_{a_1\rho\pi}\alpha)^2}{48\pi g_\rho^4m_{K^+}^3}
Y_{a_1^+}Z_{K^+} \nl
&&\times \lambda^{3/2}(m_{K^+}^2,m_{\pi^+}^2,M^2)\
\sqrt{M^2-4m_\ell^2}\left(1+\frac{2m_\ell^2}{M^2}\right)
\left(\frac{m_\rho^2}{m_\rho^2-M^2}\right)^2\ .
\eea
Besides $g_{a_1\rho\pi}$ there are two other nontrivial parameters
entering formula (\ref{kpiee}), $Y_{a_1^+}$ and $Z_{K^+}$. Their
values can be found in Tables \ref{ymtab} and \ref{fvtab}, respectively.
When we integrate (\ref{kpiee}) over the full range of dielectron
masses and use the experimental value of the $K^+$ lifetime, we get
the branching fraction $B(K^+\ra\pi^+\die)\approx 3.1\dek{-7}$. The
experimental value is $(2.74\pm0.23)\dek{-7}$. The uncertainty of
our result comes from the $a_1\rho\pi$ coupling constant, which is given
by the poorly known (and understood) width of the $a_1$ meson.
But this uncertainty disappears when we calculate the branching ratio
of the $\dim$ and $\die$ modes, which is a function only of the masses
of participating particles. In Table \ref{neutralt} we show therefore
the branching fraction of the $K^+\ra\pi^+\dim$ mode normalized by
the experimental value of the dielectron one.

As we have already indicated, there are other MD diagrams that
can contribute to the amplitude of $K^+\ra\pi^+\dil$ decay mode.
First of all, it might be a diagram obtained from that in
Fig.~\ref{kpieef} by substitutions $a_1^+\ra\rho^+$ and
$\rho^0\ra\omega,\  \phi$. But it vanishes identically as a consequence
of the presence of the totally antisymmetric Levi-Civitta tensor in the
hadronic vertex together with the pion momentum in the $W^+\pi^+$
junction.

\begin{figure}
\begin{center}
\leavevmode
\setlength \epsfxsize{15cm}
\epsffile{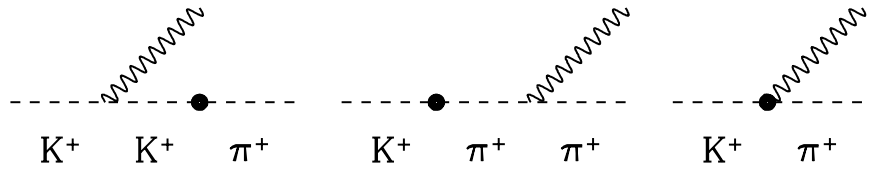}
\end{center}
\caption{Three contributions to the matrix element of the decay
$K^+\ra\pi^+\gamma (\gamma^*)$ related to Lagrangians (5.5) and (5.6).}
\label{kpigamf}
\end{figure}

Then we have diagrams that contain only pseudoscalar mesons. They are
generated by taking the basic diagram in which $K^+$ converts to
$\pi^+$ via $W^+$ and attaching a virtual photon alternatively to all
possible lines. It can be shown that the sum of those diagrams is
vanishing. To simplify the discussion we will show it here in the limit
of an infinitely heavy $W$ boson. In this limit we can introduce the
following effective Lagrangian for the  weak interaction between pions
and kaons:
\be
\label{lagpik}
{\cal L}_{\pi K}=-\frac{G_F}{\sqrt 2}f_\pi V_{ud}f_K V_{us}
\partial^\mu\varphi_\pi\partial_\mu\varphi_K^\dagger\ + \ {\rm h.c.}\ .
\ee
After switching on the electromagnetic interaction by the minimal
substitution principle, we are getting not only the usual terms
describing the emission of a photon from the pion and kaon lines,
but also contact terms generated from (\ref{lagpik}). The one-photon
part of the electromagnetic interaction Lagrangian thus reads
\be
\label{lagpikga}
{\cal L}_{\gamma}=iea_\mu\left[\varphi_\pi^\dagger\partial^\mu
\varphi_\pi+\varphi_K^\dagger\partial^\mu\varphi_K
-\frac{G_F}{\sqrt 2}f_\pi V_{ud}f_K V_{us}\left(
\varphi_\pi^\dagger\partial^\mu
\varphi_K+\varphi_K^\dagger\partial^\mu\varphi_\pi\right)\right]\ ,
\ee
where $a_\mu$ denotes the electromagnetic field operator. Now it is
easy to check that the matrix element for the photon production (both
real or virtual) calculated as the sum of the emission from kaon line,
emission from pion line, and the contact term, see Fig.~\ref{kpigamf},
is identically zero. It is a consequence of our treating pions and
kaons as elementary quanta. In \cite{vain76} this kind of contribution
was calculated assuming nontrivial electromagnetic structure of the
participating mesons. The result is proportional to the difference
between kaon and pion electromagnetic radii squared.

\begin{figure}
\begin{center}
\leavevmode
\setlength \epsfxsize{15cm}
\epsffile{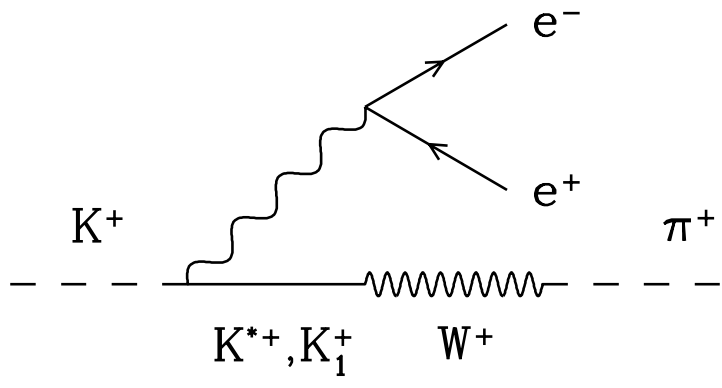}
\end{center}
\caption{Other two possible MD Feynman diagrams for
$K^+\ra\pi^+e^+e^-$.}
\label{kpieekst}
\end{figure}

The last two diagrams conceivable in the lowest order of MD are
illustrated in Fig.~\ref{kpieekst}. The matrix element with the $\ksp$
in the intermediate state vanishes identically. The contribution from
$K_1^+$ is nonvanishing but small. This can be seen from the following:
When we consider this part of the transition amplitude separately,
ignoring the contribution from  the $a_1$ diagram (Fig.~\ref{kpiee}),
the resulting branching fraction of $K^+\ra\pi^+e^+e^-$ can be expressed
in terms of the decay width of $K_1^+\ra K^+\gamma$. To get the correct
experimental number for the former, the latter had to be unrealistically
high, about 40\%.

\begin{figure}
\begin{center}
\leavevmode
\setlength \epsfxsize{15cm}
\epsffile{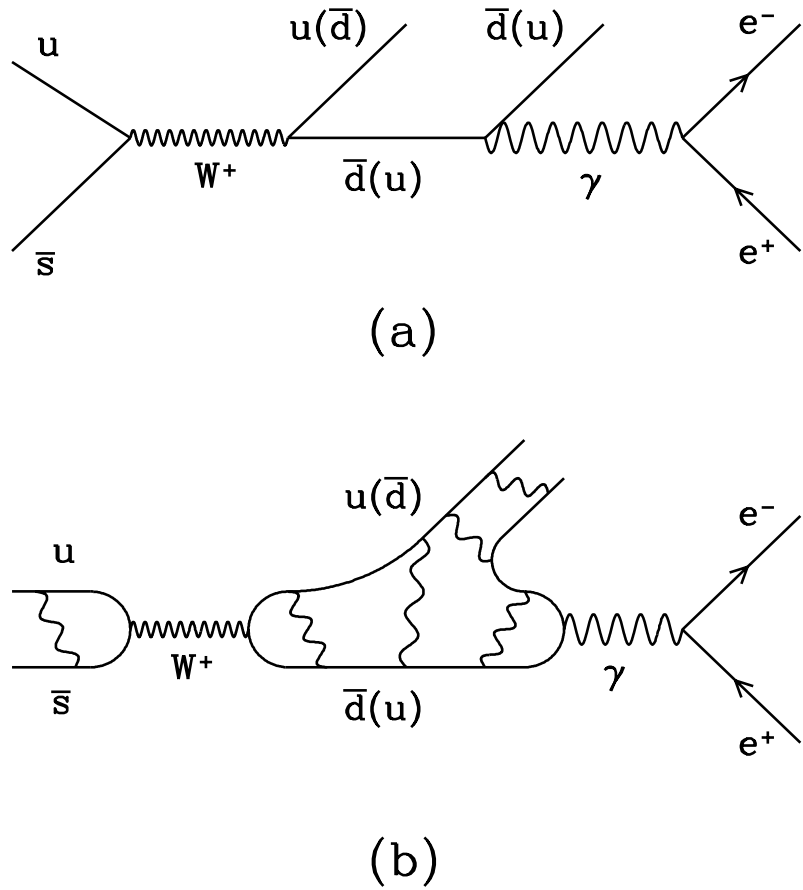}
\end{center}
\caption{Selected quark diagrams of the $K^+\ra\pi^+e^+e^-$
decay mode. (a) Two (differing by $u\leftrightarrow\bar d$) of possible
electroweak diagrams; (b) After the strong interactions are switched on,
the previous diagrams develop into those that provide the most important
contributions, like the one shown here. Unlabeled wavy curves represent
gluons.}
\label{kpieefqq}
\end{figure}

To complete our discussion about the $K^+\ra\pi^+\die$ mode let us
stress that in the MD approach we have gotten a parameter-free
description of its decay rate, dominated by the $a_1$ diagram in
Fig.~\ref{kpieef}. Other meson diagrams give smaller contributions.
Nevertheless, they will have to be taken into account when a more
detailed comparison with the next generation of more precise data
is made. Our result suggests that in any  approach
based on the standard model it is important to consider the diagram
depicted in Fig.~\ref{kpieefqq}a. It represents a seed for the class
of diagrams, like the one shown in Fig.~\ref{kpieefqq}b, into which
it develops after QCD corrections are included. This class corresponds
to the most important meson diagram, Fig.~\ref{kpieef}.

Finally, it has to be stressed that the successful description of
the decay $K^+\ra\pi^+\die$ was possible because the short-distance
part of the amplitude, which contains contributions from the
electromagnetic penguin $s\ra d+\gamma^*$, the $Z^0$ penguin
$s\ra d+Z^{0*}$, and the $W$ box diagram, is about three orders of
magnitude smaller than the long-distance part \cite{buchalla}.

To get an estimate of the branching fraction of the transition of
charmed pseudoscalar mesons $D^+$ and $D_s^+$ to a dilepton and
a pion we will again use Eq.~(\ref{kpiee}) with obvious modifications.
The results shown in Table~\ref{neutralt} should really be considered
as an order-of-magnitude estimate because (i) the $a_1$ in
Fig.~\ref{kpiee} is very far from its mass-shell and its coupling
to $W$ may differ from that assumed in (\ref{ccvavd}); (ii) also
higher charged pseudoscalar [for example, $\pi(1300)$] or tensor
[$a_2(1320)$] resonances that couple to the $\rho\pi$, $\omega\pi$,
or $\phi\pi$ system can appear in the intermediate state. But it is
highly improbable that the observed branching fractions will be
dramatically lower than those shown in Table~\ref{neutralt} as
a result of destructive interference. In fact, the matrix element
now is not a mere number, but a function of the dilepton mass and the
angle between the dilepton and pion. A substantial cancellation
would require the same functional dependence of different
contributions.

In spite of all the crudeness of our estimates we can say that the MD
approach predicts the branching fraction of $D^+\ra\pi^+\dil$ that is at
least by an order of magnitude higher than the prediction of a standard
model calculation \cite{hewett}. But even the MD prediction is about two
orders of magnitude below the present-day experimental limit. The same
is true for the MD prediction in the $D_s\ra\pi^+\dil$ case.

\begin{figure}
\begin{center}
\leavevmode
\setlength \epsfxsize{15cm}
\epsffile{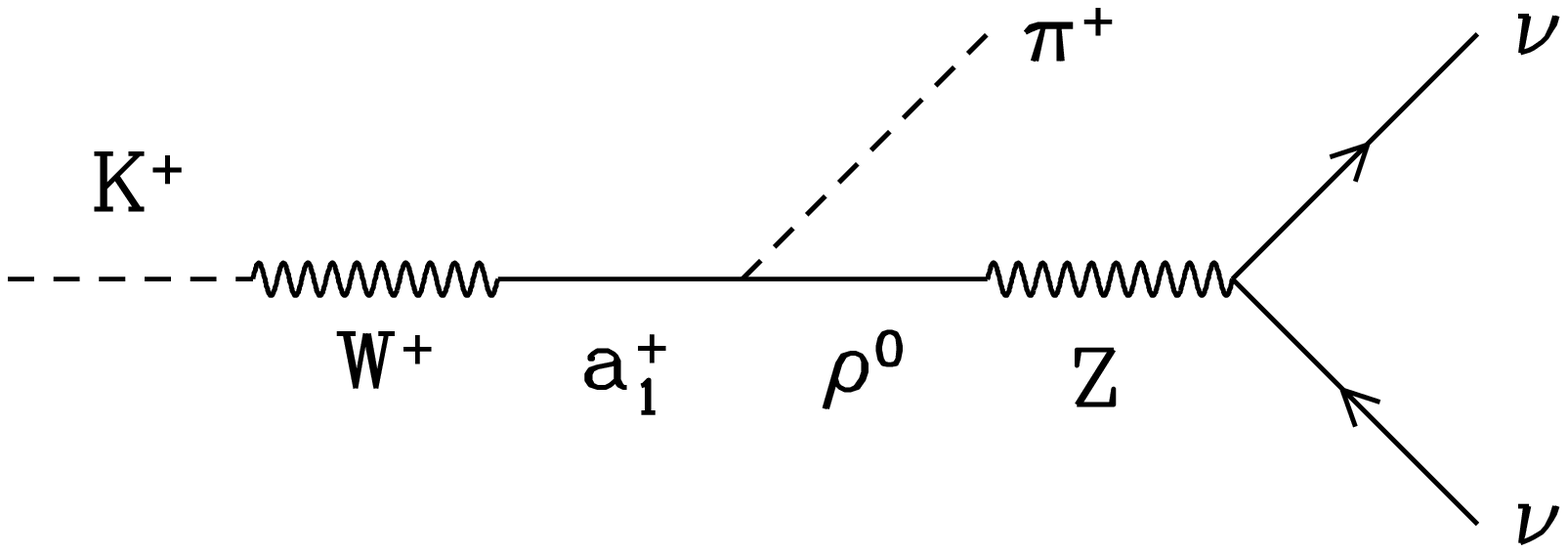}
\end{center}
\caption{Long-distance part of the matrix element of the CP conserving
decay $K^+\ra\pi^+\nu\bar{\nu}$.}
\label{kpinunuf}
\end{figure}

To check the applicability of the MD approach to neutral current
processes mediated by the neutral gauge boson $Z$ we calculated the
long-distance contribution to the CP conserving decay
$K^+\ra\pi^+\bar\nu\nu$. In the MD approach it proceeds mainly
according to diagram displayed in Fig.~\ref{kpinunuf}. Its differential
partial width is given by
\be
\label{kpinunu}
\frac{d\Gamma_{K^+\ra\pi^+\bar\nu\nu}}{dt}=\frac{G_F^4
g_{a_1\rho\pi}^2Y_{a_1^+}Z_{K^+}}{3g_\rho^4(8\pi m_{K^+})^3}
(1-2\sin^2\theta_W)^2\ t^2\ \lambda^{3/2}(m_{K^+}^2,m_{\pi^+}^2,t)
\left(\frac{m_\rho^2}{m_\rho^2-t}\right)^2,
\ee
where $t=(p_K-p_\pi)^2$ is the four-momentum transfer squared, or,
equivalently, the mass of the $\bar\nu\nu$ system. The integrated
branching fraction  $(7.9\pm0.6)\times10^{-18}$ does not have any
observational value. From the theoretical point of view it is
interesting and perhaps surprising that our value is practically
equal to the recent estimate $7.71\times10^{-18}$ (error not given)
\cite{geng} obtained from the finite part of the one-loop amplitude
in the chiral perturbation theory.

\section{Relation between the conserved vector current and
meson dominance hypotheses}
\label{cvcvermd}

The conserved vector current (CVC) hypothesis is a useful concept in
the weak interaction phenomenology. From a pragmatic point of view it
enables the decay rate of some flavor conserving weak processes to be
related to the data on hadron production in $\die$ annihilation
\cite{okun,tsai,gilman,eidelman}. It is therefore natural to ask what
is the relation between MD and CVC, which of the two approaches is more
general, and which has more predictive power.

It is evident that MD can be applied also to processes during which
the flavor of the hadronic system changes, whereas the CVC cannot be.
Let us therefore consider only the flavor conserving processes with
the weak vector current. At first sight it seems that MD in weak
interactions is a straightforward consequence of CVC hypothesis and
$\rho^0$ dominance in electromagnetic interactions. If it were true,
the two concepts would lead to the same results in the region where
their domains of validity overlap.

A typical process of this type is the $\pi_{e3}$ decay
$\pi^+\ra\pi^0 e^+\nu_e$. We have shown that one can get good
agreement with the data by calculating its branching fraction from the
MD Lagrangian (\ref{ccvavd}) without any further assumption. It has also
been claimed for a long time (see, e.g., \cite{okun,pietschm}) that the
agreement of the CVC result with experiment is perfect and lends
strongest support to the CVC hypothesis. To examine this assertion let
us sketch briefly the central point of the CVC derivation.

From general principles it follows that the matrix element for the
$\pi_{e3}$ decay has the form (\ref{ffdef}). The continuity equation
for the conserved weak vector current requires
\be
\label{cvc}
(p_1-p_2)_\mu\left[(p_1+p_2)^\mu f_+(t)+(p_1-p_2)^\mu f_-(t)\right]=0\ ,
\ee
what leads to the relation
\be
\label{cvcff}
f_-(t)=-\frac{m_{\pi^+}^2-m_{\pi^0}^2}{t}\ f_+(t)\ .
\ee
Because $t$ is the mass squared of the $\ell\nu$ system, it cannot vanish.
In the limit of exact isospin symmetry we have $m_{\pi^+}=m_{\pi_0}$ and
therefore
\be
\label{zerofm}
f_-(t)\equiv 0\ .
\ee
The usual CVC result for the $\pi_{e3}$ branching fraction is obtained
by assuming that the identity (\ref{zerofm}) holds also when isospin
symmetry is broken. The other assumptions state that the function
$f_+(t)$ can be replaced by a constant in the small $t$ range allowed
kinematically in the $\pi_{e3}$ decay and that the relations among
different components of the electroweak isovector vector current
remain same as in the case of exact isospin symmetry. The final formula
can be found in Ref.~\cite{pietschm}.\footnote{In Ref.~\cite{okun}
additional approximations were made, which lowered the result by
$2.5\sigma$. Equation~(7.15) in \cite{pietschm} contains an obvious
misprint: $\pi^5$ should be read as $\pi^3$.} Although different from the
MD formula, after numerical evaluation it gives practically identical
value, which agrees with the experimental branching ratio very well
(see Table~\ref{pspslnut}).

But let us look at the CVC procedure described above a little more
closely. The assumption that relation (\ref{zerofm}) is valid also when
isospin symmetry is broken violates the relation (\ref{cvc}). So the
usual CVC result is, in fact, obtained not by assuming the conservation
of the vector current, but rather assuming a special type of its
nonconservation, namely such that results in Eq. (\ref{zerofm}).

If we strictly enforce the conservation of the vector current by
honoring Eq.~(\ref{cvcff}), which follows from it, we obtain\footnote{
For the differential decay rate formula see Appendix~\ref{genff}.}
the $\pi_{e3}$ branching fraction of $(0.8872\pm0.0019)\dek{-8}$,
which disagrees with the contemporary experimental value
$(1.025\pm0.034)\dek{-8}$.

To conclude: We found a process for which the CVC hypothesis and MD
give different results. Meson dominance in the flavor conserving
vector current sector thus represents a dynamic assumption that
is different from what would be obtained by merging the CVC hypothesis
with the VMD in electromagnetic interactions. The case of the
$\pi_{e3}$ decay suggests that MD is better suited for description
of the processes in which the isospin invariance is broken.

\section{Conclusions and comments}
\label{comments}

The hypothesis that the weak interaction of hadronic systems
at low energies is dominated by the coupling of the vector,
axial-vector, and pseudoscalar mesons to the gauge bosons has
been scrutinized.  The strength of the weak coupling of the $\rho(770)$
meson is uniquely determined by vector-meson dominance in electromagnetic
interactions; flavor and chiral symmetry breaking effects modify the
coupling of other vector mesons and axial-vector mesons. Corresponding
strengths parameter and their products with (mostly unknown) strong
interaction coupling constants constitute the free parameters of
our approach. They are fixed by experimental data on the branching
fractions of the selected decay modes of the $\tau$ lepton and
semileptonic decay modes of pseudoscalar mesons. Some hadronic coupling
constants were determined from the widths of strong and radiative decays.

After fixing the parameters, many decay rates of the $\tau$
lepton and pseudoscalar mesons ($\pi$, $K$, $D$, $D_s$, $B$, and $B_s$)
have been calculated and compared to experimental data. They fall into
three categories:

\begin{itemize}
\item
 Decay modes where the calculated result is in good agreement with
    observation. One can expect that these modes, when calculated in the
    framework of the standard model, are dominated by the weak
    quark-antiquark annihilation and creation diagrams. The nicest
    example in this category is the semileptonic decay
    $\pi^+\ra\pi^0 e^+\nu_e$. The calculated branching fraction
    is $(1.0041\pm0.0021)\times10^{-8}$, while experiment says
    $(1.025\pm0.034)\times10^{-8}$. Many nonleptonic decays are
    also well described. For example, $D^+\ra\pi^0\pi^+$,
    $D_s^+\ra\eta\pi^+$, $B^+\ra\bar D^0D_s^+$,
    $B^0\ra D^-\rho^+$. Also the branching fraction of a quite complex
     mode $\tau^-\ra\eta\pi^-\pi^0\nu_\tau$ agrees nicely with the
    experimental figure. So does that of the ``neutral current flavor
    changing" mode $K^+\ra\pi^+ e^+e^-$.
\item
 Decay modes where the calculation disagrees with experimental data.
     Here the standard model diagrams that do not have an analogy
     in the meson dominance approach ($W$ emission or absorption from
     a quark line, penguin diagrams, box diagrams, etc.) are expected
     to dominate. The two pion decays of the $K$ mesons are a typical
     example.
\item
 Decay modes that have not been measured yet. Some meson-dominance
      predictions: $B[K_L^0\ra K^\pm e^\mp\bar\nu_e(\nu_e)]=
      (3.4\pm0.6)\times10^{-9}$, $B(B^0\ra\pi^-\tau^+\nu_\tau)=
      (9.4\pm3.1)\times10^{-5}$, $B(B_s^0\ra D_s^-\tau^+\nu_\tau)=
      (1.9\pm0.6)\%$, $B(D_s^+\ra\eta^\prime K^+)= (1.3\pm0.5)
       \times10^{-3}$, $B(K^+\ra\pi^+\mu^+\mu^-)=(6.2\pm0.5)
      \times10^{-8}$.
\end{itemize}

An upper limit on the presence of the second-class vector current
was obtained using the experimental limit on the
$\tau^-\ra\pi^-\eta\nu_\tau$ branching fraction. In terms of the
scalar decay constant of the $a_0^-$ meson it reads $f_{a_0^-}<
7$~MeV. The upper bound is about twenty times smaller that the decay
constant of the $\pi^+$ meson.

What comes as a surprise is the ability of the MD approach to provide
a parameter-free description of the flavor changing $\Delta Q=0$
process $K^+\ra\pi^+\die$. Also the MD tree diagram calculation of the
long-distance part of the flavor changing neutral current decay
$K^+\ra\pi^+\nu\bar\nu$  gives the same result as a one-loop evaluation
in the chiral perturbation theory with certain prescription for handling
the divergent part.

The cross sections of several not yet observed reactions of
pions, kaons, and electron antineutrinos with target electrons come as
predictions of the meson dominance approach. These include, e.g.,
$\pi^+ e^-\ra \pi^0 \nu_e$, $K^+ e^-\ra \pi^0 \nu_e$,
$\bar\nu_e e^-\ra \pi^-\pi^0$.

The transitions $B\ra\bar{D}$ and $B_s\ra D_s^-$ where
the final state meson is accompanied by an $\ell^+\nu_\ell$ system,
or a positively charged pseudoscalar, vector, or axial-vector meson
cannot be explained within the MD approach without assuming the
existence of the as yet unobserved vector meson $\bcsp$. The
results of MD calculations depend on its mass. We used the value
obtained from the potential models \cite{potent1,potent2}. The question
arises whether it would not be possible to determine the $\bcsp$ mass
from the experimental branching ratios of various decay modes using
the MD formulas. In order to answer this question we increased the
$\bcsp$ mass by 0.5 GeV and recalculated the branching fractions.
The biggest decrease was experienced by semileptonic decay modes. But
even here it was only by 3.3\%. It makes any effort to predict the
$\bcsp$ mass using the MD approach unrealistic. Prospects of producing
the $\bcsp$ mesons were assessed already in early papers, e.g.,
\cite{potent1,potent2}. The present state of art can be found in
\cite{bcstar} and in references therein. There is a hope that an
observable number of $B_c$ and $B_c^*$ events can be produced at
LEP and at Tevatron.

On the theoretical side, the relation between the meson dominance and
the conserved vector current hypothesis has been clarified.

\acknowledgements

I am indebted to P.J.~Ellis, J.I.~Kapusta, A.Z.~Dubni\v{c}kov\'a,
M.~Moj\v{z}i\v{s}, and  J.~Pi\v{s}\'ut  for stimulating discussions,
to J.A.~Thompson for information about the E865 experimental program
at AGS and for discussions about the physics at $\phi$ factories.
Critical remarks by S.~Rudaz and A.~Vainshtein during the seminar in
Minneapolis in September 1995 and those of V.~\v{C}ern\'y (Bratislava,
August 1996) forced me to formulate the basic assumptions and virtues
of this approach more clearly. This work was initiated during
my stay at the University of Minnesota, which was supported by the U.S.
Department of Energy under Contract No. DOE/DE-FG02-87ER-40328
and completed at the Department of Energy's Institute for Nuclear Theory
at the University of Washington. The partial support provided by the
latter is gratefully acknowledged. In Bratislava, the work was a part
of the project VEGA 1/1323/96.

\appendix
\section{Determining the $VVP$ coupling constants from data on strong
and radiative decays}
\label{detccvvp}

If the decay $V_1\ra V_2+P$ is energetically allowed, the $V_1V_2P$
coupling constant can be determined from its empirical decay rate.
Using Lagrangian (\ref{lagvvps}) we easily derive the formula
\be
\label{gamv1v2p}
\Gamma_{V_1\ra V_2+P}=\frac{g^2_{V_1V_2P}}{96\pi m_{V_1}^3}\
\lambda^{3/2}(m_{V_1}^2,m_{V_2}^2,m_P^2)\ .
\ee
Triangle function $\lambda$ is defined by Eq. (\ref{triangle}).
The experimentally given branching fraction of $\phi\ra\rho\pi$
includes three different final states, which would be equally probable
if the isospin symmetry was exact. If we assume that the latter is
violated only through mass differences, we obtain
$g^2_{\phi\rho^+\pi^-}=(1.17\pm0.07)$~GeV$^{-2}$.

To get the expression for the rate of radiative decay $V_1\ra\gamma+P$
we only need to replace $V_2$ by $\gamma$.
\be
\label{gamv1gap}
\Gamma_{V_1\ra\gamma+P}=\frac{g^2_{V_1\gamma P}}{96\pi m_{V_1}^3}\
(m_{V_1}^2-m_P^2)^3\ .
\ee
VMD in electromagnetic interactions enables to express the
$g_{V_1\gamma P}$ coupling constant by means of the hadronic ones.
For radiative decay $\omega\ra\gamma\pi^0$ the situation is simple
because only $\rho^0$ can couple to the $\omega\pi^0$ system. Using
(\ref{vgamma}) we can write
\[
g^2_{\omega\gamma\pi^0}=\frac{4\pi\alpha}{g_\rho^2}
\ g^2_{\omega\rho^0\pi^0}
\]
and calculate
\be
\left(\frac{g_{\omega\rho^0\pi^0}}{g_\rho}\right)^2=
\frac{24m_\omega^3}{\alpha
\left(m_\omega^2-m_{\pi^0}^2\right)^3}\Gamma_{\omega\ra\gamma\pi^0}=
(5.40\pm0.32){\rm GeV}^{-2}
\ee

The radiative decay $\rho^0\ra\eta\gamma$ can proceed only via the
strong $\rho^0\eta\rho^0$ vertex because the isospin conservation
prevents coupling of the $\rho^0\eta$ system to either $\omega$ or
$\phi$. Another consequence of the isoscalar character of the $\eta$
meson is that the quantity
\be
\left(\frac{g_{\rho\eta\rho}}{g_\rho}\right)^2=\frac{24m_\rho^3}{\alpha
\left(m_\rho^2-m_\eta^2\right)^3}\Gamma_{\rho^0\ra\eta\gamma}=
(15.1\pm2.8){\rm GeV}^{-2}
\ee
has the same value for all charge states of $\rho$.

The case of $K^*$ radiative decays is most complicated because the
resulting amplitude is given as a coherent sum of three amplitudes with
$\gamma$ coupled to $\rho$, $\omega$, and $\phi$. In spite of this
complication we can determine the $K^*K\rho$ coupling constants because
the $\omega$ and $\phi$ contributions to $\ksp\ra K^+\gamma$ are
equal to those to $K^{*0}\ra K^0\gamma$, whereas the $\rho^0$
contribution changes sign. We are thus getting the set of equations
\bea
(x+y)^2&=&a_+\nl
(-x+y)^2&=&a_0 \ , \nonumber
\eea
where $x=g_{\ksp K^+\rho^0}/g_\rho$, $y$ stands for the expression that
contains only isoscalar coupling constants, and
\[
a_c=\frac{24m_{K^{*c}}^3}{\alpha\left(m_{K^{*c}}^2-m_{K^c}^2\right)^3}
\Gamma_{K^{*c}\ra K^c+\gamma}
\]
for $c=+,0$,  Using the relation
$g_{\ksp K^0\rho^+}^2=2g_{\ksp K^+\rho^0}^2$, which follows from isospin
invariance, we eventually get two values of $(g_{\ksp K^0\rho^+}
/g_\rho)^2$ that are compatible with experimental data on radiative
$K^*$ decays, namely $2.21\pm0.14~{\rm GeV}^{-2}$ and $(9.2\pm2.9)
\dek{-2}$~GeV$^{-2}$.

\section{Decay \protect{$P_1\ra P_2+\ell^+\nu_\ell$}: General form
factors}
\label{genff}

The general form of the matrix element is
\[
{\cal M}= C\ \left[f_+(t)(p_1+p_2)^\mu+f_-(t)(p_1-p_2)^\mu\right]
\bar\ell\gamma_\mu(1-\gamma_5)\nu \ ,
\]
where $p_1$ ($p_2$) is the four-momentum of the incoming (outgoing)
pseudoscalar meson and $t=(p_1-p_2)^2$ is the mass squared of the
lepton system. Using (\ref{3body}) and integrating over the solid
angle in the $\ell^+\nu_\ell$ rest frame we get the following
expression for the differential decay rate:
\bea
\label{dpplnugf}
\frac{d\Gamma_{P_1\ra P_2\ell^+\nu_\ell}}{dt}&=&\frac{|C|^2}
{3(4\pi m_{P_1})^3}\ \frac{t-m_\ell^2}{t^3}\ \lambda^{1/2}
(m_{P_1}^2,m_{P_2}^2,t)
\left\{\varphi_1(t)|f_+(t)|^2\right.\nl
&&\left. +6zt(x-y)(t-z){\rm Re}[f^*_+(t)f_-(t)]
+3zt^2(t-z)|f_-(t)|^2\right\}.
\eea
Function $\varphi_1(t)$ is defined in Eq. (\ref{phi1}). Also the meaning
of other symbols is same as in Sec.~\ref{pspslnu}: $x=m_{P_1}^2$,
$y=m_{P_2}^2$, $z=m_\ell^2$.

To get the total rate of $\pi^+\ra\pi^0 e^+\nu_e$ that follows from
the requirement of exact conservation of the vector current,
see Section~\ref{cvcvermd}, we need
to integrate (\ref{dpplnugf}) with $C=G_FV_{ud}$, $f_+(t)=1$, and
$f_-(t)=-(x-y)/t$.

\bt
\caption{Parameters
$Y_V$ and $Y_A$, defined by Eq.~(2.16) and
characterizing the coupling of vector and axial-vector mesons to
the charged gauge bosons, their sources, and values of parameters
$w$ extracted from them.}
\label{ymtab}
\btab{lccl}
V, A & $Y_V$, $Y_A$ & $w_V$, $w_A$ & Source \\
\hline
$\rho^+$ & $0.9479\pm0.0019$ & 1 & $\Vud^2$ \\
$\ksp$   & $(4.20\pm0.09)\dek{-2}$  & $ 0.929\pm0.013 $ &
$K^+\ra\pi^0e^+\nu_e$, $K^*\ra K\pi$ \\
$\dssp$ & $0.7\pm0.5$ & $0.83\pm0.33$ & $B^0\ra D^-\dssp$\\
$a_1^+$ & $0.6134\pm0.0032$ & $0.8044\pm0.0023$ &
$\tau^-\ra a_1^-\nu_\tau$\\
$K_1^+$ & $(3.4\pm1.7)\dek{-2}$ & $0.84\pm0.21$ &
$\tau^-\ra K_1^-\nu_\tau$
\etab
\et

\bt
\caption{Parameters $Z_{P}$ characterizing the coupling
of pseudoscalar mesons to the charged gauge boson and their sources.
For definition, see Eq.~(2.20).}
\label{fvtab}
\btab{lcl}
P & $Z_{P}$ (GeV$^2$) & Source \\
\hline
$\pi^+$ & $(1.6419\pm0.0010)\dek{-2}$ &
$\pi^+\ra \mu^+\nu_\mu\ + \ \mu^+\nu_\mu\gamma$\\
$K^+$ & $(1.247\pm0.004)\dek{-3}$ &
$K^+\ra \mu^+\nu_\mu\ + \ \mu^+\nu_\mu\gamma$\\
$D^+$ & $(2.2\pm0.5)\dek{-3}$& $f_{D^+}$ from \cite{soni}, $\Vcd$ from
\cite{pdg}\\
$\dsp$ & $(1.1\pm0.5)\dek{-1}$ & $\dsp\ra\mu^+\nu_\mu$
\etab
\et

\bt
\caption{Branching fractions of the $\tau$ lepton calculated in the
MD approach and comparison with experimental data. Column
C shows the meson coupled to the weak gauge boson.}
\label{tautabl}
\btab{llccl}
Final state & C & Meson Dominance result& Data & Notes\\
\hline

$\pi^-\nu_\tau$ & $\pi^-$ &
$(10.91\pm0.06)\%$ & $(11.31\pm0.15)\%$ &  \\

$ K^-\nu_\tau$ & $ K^-$ &
$(7.13\pm0.04)\dek{-3}$ & $(7.1\pm0.5)\dek{-3}$ &  \\

$\pi^-\pi^0\nu_\tau$ & $\rho^-$ &
$(24.4\pm0.4)\%$ & $(25.24\pm0.16)\%$ & (a) \\

$K^-K^0\nu_\tau$ & $\rho^-$ & & $(1.55\pm0.28)\dek{-3}$ & (b) \\

$\pi^-\omega\ \nu_\tau$ & $\rho^-$ &
$(1.22\pm0.56)\%$ & $(1.84\pm0.05\pm0.14)\%$ & (c,d) \\

$\pi^-\phi\ \nu_\tau$ & $\rho^-$ &$(1.20\pm0.48)\dek{-5}$&
$<3.5\dek{-4}$  & (c) \\

$\eta\ \pi^-\pi^0\nu_\tau$ & $\rho^-$ & $(1.79\pm0.33)\dek{-3}$
& $(1.71\pm0.28)\dek{-3}$ & (a) \\

$K^*(892)^-\nu_\tau$ & $K^{*-}$ &$(1.06\pm0.03)\%$
& $(1.28\pm0.08)\%$ & (e)\\

$ a_1^-\nu_\tau$ & $a_1^-$ & & $(18.11\pm0.37)\%$ & (d,e,f) \\

$K_1(1400)^-\nu_\tau$&$K_1^-$&&$(8\pm4)\dek{-3}$& (e,g)\\

$\eta\ \pi^-\nu_\tau$ & $a_0^-$ &   & $< 1.4\dek{-4}$ & (h)
\etab
(a) The normalization is determined by the VMD in QED. (b) Used to fix
$X_{\rho^-K^-K^0}=0.64\pm0.12$,
which differs from what one would get from the $SU_f(4)$ coupling
constant ratio by about 20\%.
(c) Calculation by Castro and Falc\'{o}n \cite{castro}.
(d) Experimental value taken from \cite{buskulic}.
(e) MD calculation in the narrow width approximation.
(f) Used to fix $Y_{a_1}$.
(g) Used to fix $Y_{K_1}$.
(h) Coupling of the $\pi^-\eta$ system to $\pi^-$, $\rho^-$, or $a_1^-$
is forbidden by the strong interaction and spin-parity conservation laws.
This mode put a limit on the $a_0^-(980)$ decay constant
$f_{a_0^-}<7$~MeV.
\et

\bt
\caption{Parameters $X_{P_1P_2V}$, defined by Eq.~(4.1): Numerical
values and their sources.}
\label{xtab}
\btab{ccccll}
 $P_1$ & $P_2$ & $V$ &   $X_{P_1P_2V}$ & Source & Notes\\
\hline
$\pi^+$&$ \pi^0$&$\rho^+$&$(0.9479\pm0.0020)$     &$\Vud^2 $\\

$K^+  $& $\pi^0$ & $\ksp  $ & $(1.206\pm0.015)\dek{-2}$ &
$K^+ \ra \pi^0 e^+ \nu $   \\

$K^0 $& $\pi^-$&$\ksp  $&$(2.412\pm0.030)\dek{-2}$&
$2\times X_{K^+\pi^0\ksp}$ \\

$D^+$&$ \pi^0$&$D^{*+}$  &$(8.9\pm3.4)\dek{-3}$&
$D^+ \ra \pi^0 \ell^+ \nu$ \\

$D^0$&$ \pi^-$&$D^{*+}$ &$(1.8\pm0.7)\dek{-3}$ &$2\times
X_{D^+\pi^0D^{*+}}$ \\

$D^+$&$ \ak  $&$\dssp $&$0.263\pm0.015$&$= X_{D^0K^-\dssp} $ \\

$D^0$& $K^-  $&$\dssp$ &$0.263\pm0.015$&$D^0 \ra K^-  \mu^+ \nu$ \\

$\dsp$&$\eta  $&$\dssp$  &$0.139\pm0.039$  &
$\dsp \ra \eta \ell^+ \nu$ \\

$\dsp$&$\eta^\prime$&$\dssp$&$0.18\pm0.07$ &
$\dsp\ra\eta^\prime\ell^+\nu$\\

$B^+$ & $\pi^0$ &$ B^{*+} $& $(4.3\pm1.4)\dek{-7}$ &
$=X_{B^+\pi^-B^{*+}}/2$\\

$B^0$ & $\pi^-$ & $B^{*+}$ & $(8.5\pm2.8)\dek{-7}$ &
$\bar B^0\ra\pi^-\ell^+\bar\nu_\ell $\\

$B^+$&$\ad   $&$\bcsp$ &$(3.5\pm0.9)\dek{-4}$&
$B^0 \ra D^- \ell^+ \nu$&(a)\\

$B^0$&$D^-   $&$\bcsp$ &$(3.5\pm0.9)\dek{-4}$&$=X_{B^+D^-\bcsp}$&
(a) \\

$B_s^0$&$D_s^-$&$\bcsp$ &$(1.3\pm0.4)\dek{-3}$&
$B_s^0 \ra D_s^-\ell^+ \nu$&(a) \\

$\ak $&$ K^-  $&$\rho^+$ &$(0.64\pm0.12)$&$\tau^-\ra K^-K^0\nu_\tau$

\etab
(a) Existence of $\bcsp$ with a mass of 6.34 GeV/$c^2$ assumed.
\et

\bt
\caption{Branching fractions of semileptonic $P_1 \ra P_2$
transitions calculated in the MD approach and comparison with experimental
data. Column C shows the meson coupled to the weak gauge boson.}
\label{pspslnut}
\btab{llccl}
Decay mode& C & MD result & Data & Notes\\
\hline

$\pi^+\ra\pi^0e^+\nu_e$ & $\rho^+$ &
$(1.0041\pm0.0021)\dek{-8}$ &
$(1.025\pm0.034)\dek{-8}$ & (a) \\

$K^0_L\ra K^\pm e^\mp\bar\nu_e(\nu_e)$
& $\rho^\mp$ &
$(3.4\pm0.6)\dek{-9}$ & & (b,c,d) \\

$K^+\ra\pi^0e^+\nu_e$ & $K^{*+}$ &   &
$(4.82\pm0.06)\%$ & (e)\\

$K^+\ra\pi^0\mu^+\nu_\mu$ & $K^{*+}$ &
$(3.10\pm0.04)\%$ & $(3.18\pm0.08)\%$ \\

$K^0_L\ra\pi^\pm e^\mp\bar\nu_e(\nu_e)$
& $K^{*\mp}$ &
$(40.7\pm0.5)\%$ & $(38.78\pm0.27)\%$ & (c) \\

$K^0_L\ra\pi^\pm\mu^\mp\bar\nu_\mu(\nu_\mu)$
& $K^{*\mp}$ &
$(26.18\pm0.33)\%$ & $(27.17\pm0.25)\%$ & (c)\\

$K^0_S\ra\pi^\pm e^\mp\bar\nu_e(\nu_e)$
& $K^{*\mp}$ &
$(7.03\pm0.12)\dek{-4}$ & $(6.70\pm0.07)\dek{-4}$ & (c,f) \\

$K^0_S\ra\pi^\pm\mu^\mp\bar\nu_\mu(\nu_\mu)$
& $K^{*\mp}$ &
$(4.52\pm0.08)\dek{-4}$ & $(4.69\pm0.06)\dek{-4}$ & (c,f)\\

$D^+\ra\pi^0\ell^+\nu_\ell$ & $D^{*+}$ &
& $(5.7\pm2.2)\dek{-3}$ & (e,g)\\

$D^0\ra\pi^-\ell^+\nu_\ell$ & $D^{*+}$ & $(4.4\pm1.7)\dek{-3}$
& $(3.8^{+1.2}_{-1.0})\dek{-3}$ & (g)\\

$D^0\ra K^-\mu^+\nu_\mu$ & $D_s^{*+}$ & & $(3.23\pm 0.19)\%$ & (e)\\

$D^0\ra K^-e^+\nu_e$ & $D_s^{*+}$ & $(3.33\pm0.20)\%$
& $(3.64\pm 0.20)\%$ &  \\

$D^+\ra \overline K^0\mu^+\nu_\mu$ & $D_s^{*+}$ &$(8.3\pm0.5)\%$
& $(7.0^{+3.0}_{-2.0})\%$ & \\

$D^+\ra \overline K^0 e^+\nu_e$ & $D_s^{*+}$ &$(8.6\pm0.5)\%$
& $(6.6\pm0.9)\%$ & \\

$D^+_s\ra \eta \ell^+\nu_\ell$ & $D_s^{*+}$ &
& $(2.5\pm0.7)\%$ & (e,g)\\

$D^+_s\ra \eta^\prime \ell^+\nu_\ell$ & $D_s^{*+}$ &
& $(8.7\pm3.4)\dek{-3}$ & (e,g)\\

$B^+\ra\pi^0 e^+\nu_e$ & $B^{*+}$ & $(9.4\pm3.1)\dek{-5}$
& $<2.2\dek{-3}$ & \\

$B^+\ra\pi^0\tau^+\nu_\tau$ & $B^{*+}$ & $(4.9\pm1.6)\dek{-5}$
&  &(b) \\

$B^0\ra\pi^-\ell^+\nu_\ell$ & $B^{*+}$ &
& $(1.8\pm0.6)\dek{-4}$ & (g,h)\\

$B^0\ra\pi^-\tau^+\nu_\tau$ & $B^{*+}$ & $(9.4\pm3.1)\dek{-5}$
&  &(b) \\

$B^0\ra D^- \ell^+\nu_\ell$ & $B_c^{*+}$ &
& $(1.9\pm0.5)\%$ & (e,g,i)\\

$B^0\ra D^- \tau^+\nu_\tau$ & $B_c^{*+}$ &
$(4.7\pm1.2)\dek{-3}$& &(b,i) \\

$B^+\ra \overline D^0 \ell^+\nu_\ell$ & $B_c^{*+}$ &
$(2.0\pm0.5)\%$ & $(1.6\pm0.7)\%$ &(g,i) \\

$B^+\ra \overline D^0 \tau^+\nu_\tau$ & $B_c^{*+}$ &
$(4.9\pm1.3)\dek{-3}$ &  & (b,i)\\

$B^0_s\ra D_s^- \ell^+\nu_\ell$ & $B_c^{*+}$ &
& $(7.6\pm2.4)\%$ & (e,g,i)\\

$B^0_s\ra D_s^- \tau^+\nu_\tau$ & $B_c^{*+}$
& $(1.9\pm0.6)\%$ & &(b,i)

\etab
(a) Using $\Vud=0.9736\pm0.0010$.
(b) Not measured yet.
(c) The sum of the charge states indicated.
(d) Hadronic coupling constant fixed by $\tau^- \ra K^-K^0\nu_\tau$.
(e) Used to fix normalization.
(f) Experimental value was calculated from $\kl$ semileptonic rate
    and the $\ks$ lifetime assuming $\Delta S=\Delta Q$ \cite{pdg}.
(g) Average of the $e^+$ and $\mu^+$ branching fractions.
(h) Used to determine $X_{B^0\pi^-B^{*+}}$. The experimental value
    is $(1.8\pm0.4\pm0.3\pm0.2)\dek{-4}$ \cite{cleojuly}, where errors
    are statistical, systematic, end estimated model dependence.
    We took the liberty of summing the errors quadratically.
(i) The existence of $B_c^{*+}$ assumed with a mass of 6.34 GeV/$c^2$.
\et

\bt
\caption{Branching fractions of the $P_1\ra P_2+P_3$ decay modes
in the tree level of MD and comparison with experimental
data. Column C shows the meson coupled to the $W^+$ boson.
Only results that do not contradict existing data are listed.}
\label{pspspstg}
\btab{llccl}
Decay mode& C & Meson dominance result & Data & Notes\\
\hline

$D^+\ra\pi^0\pi^+$ & $D^{*+}$ & $(4.0\pm1.6)\dek{-3}$ &
$(2.5\pm0.7)\dek{-3}$ &  \\

$D^0\ra\pi^-K^+$ & $D^{*+}$ & $(2.2\pm0.9)\dek{-4}$ &
$(2.9\pm1.4)\dek{-4}$ & (a) \\

$D^+\ra\ak K^+$ & $\dssp$ & $(6.8\pm0.4)\dek{-3}$ &
$(7.2\pm1.2)\dek{-3}$ &  \\

$D^0\ra K^-\pi^+$ & $\dssp$ & $(3.8\pm0.2)\%$ &
$(3.83\pm0.12)\%$ &  \\

$\dsp\ra\eta\pi^+$ & $\dssp$ & $(2.6\pm0.7)\%$ &
$(2.0\pm0.6)\%$ &  \\

$B^+\ra\pi^0\pi^+$ & $B^{*+}$ & $(6.8\pm2.3)\dek{-6}$ &
$<1.7\dek{-5}$\\

$B^+\ra\pi^0 K^+$ & $B^{*+}$ & $(5.1\pm1.7)\dek{-7}$ &
$<1.4\dek{-5}$&(a)\\

$B^+\ra\pi^0 D_s^+$ & $B^{*+}$ & $(3.8\pm2.1)\dek{-5}$ &
$<2.0\dek{-4}$\\

$B^0\ra\pi^-\pi^+$ & $B^{*+}$ & $(1.3\pm0.4)\dek{-5}$ &
$<2.0\dek{-5}$\\

$B^0\ra\pi^-K^+$ & $B^{*+}$ & $(9.9\pm3.3)\dek{-7}$ &
$<1.7\dek{-5}$&(a)\\

$B^0\ra\pi^-D_s^+$ & $B^{*+}$ & $(7.4\pm4.1)\dek{-5}$ &
$<2.8\dek{-4}$\\

$B^+\ra\ad\dsp$ & $\bcsp$ & $(1.9\pm1.0)\%$ &
$(1.7\pm0.6)\%$ & (b) \\

$B^0\ra D^-\pi^+$ & $\bcsp$ & $(3.6\pm0.9)\dek{-3}$ &
$(3.0\pm0.4)\dek{-3}$ & (b) \\

$B^0\ra D^-\dsp$ & $\bcsp$ & $(18\pm10)\dek{-3}$ &
$(7\pm4)\dek{-3}$ & (b) \\

$B_s^0\ra D_s^-\pi^+$ & $\bcsp$ & $(1.4\pm0.5)\%$ &
$<12\%$ & (b)

\etab
(a) Doubly Cabibbo suppressed mode.
(b) Existence of $\bcsp$ with a mass of 6.34 GeV/$c^2$ assumed.
\et

\bt
\caption{Branching fractions of the $P_1\ra P_2+P_3$ decay modes
in the tree level of MD and comparison with experimental
data. Column C shows the meson coupled to the $W^+$ boson.
Only results that contradict data are listed.}
\label{pspspstb}
\btab{llccl}
Decay mode& C & Meson dominance result & Data & Notes\\
\hline

$K^+\ra\pi^0\pi^+$ & $K^{*+}$ & $(86.3\pm1.1)\%$ &
$(21.16\pm0.14)\%$ &  \\

$\ks\ra\pi^-\pi^+$ & $K^{*+}$ & $(2.52\pm0.03)\%$ &
$(68.61\pm0.28)\%$ &  \\

$\ks\ra\pi^0\pi^0$ & none & 0 & $(31.39\pm0.28)\%$ &  \\

$D^0\ra\pi^-\pi^+$ & $D^{*+}$ & $(3.1\pm1.2)\dek{-3}$ &
$(1.52\pm0.11)\dek{-3}$ &  \\

$D^0\ra\pi^0\pi^0$ & none & 0 &$(8.4\pm2.2)\dek{-4}$  &  \\

$D^+\ra\ak\pi^+$ & $\dssp$ & $(9.8\pm0.6)\%$ &
$(2.74\pm0.29)\%$ &  \\

$D^0\ra\ak\pi^0$ & none &0 & $(2.11\pm0.21)\%$ &  \\

$D^0\ra K^- K^+$ & $\dssp$ & $(2.66\pm0.16)\dek{-3}$ &
$(4.33\pm0.27)\dek{-3}$ &  \\

$D^0\ra\ak K^0$ & none & 0 & $(1.3\pm0.4)\dek{-3}$ &  \\

$\dsp\ra\eta^\prime\pi^+$ & $\dssp$ & $(1.9\pm0.8)\%$ &
$(4.9\pm1.8)\%$ &  \\

$B^+\ra\ad\pi^+$ & $\bcsp$ & $(3.7\pm1.0)\dek{-3}$ &
$(5.3\pm0.5)\dek{-3}$ & (a)

\etab
(a) Existence of $\bcsp$ with a mass of 6.34 GeV/$c^2$ assumed.
\et

\bt
\caption{Branching fractions in the tree level of MD
of the $P_1\ra P_2+P_3$ decay modes that have not yet been observed.
Only the modes with branching fractions greater than $\dek{-4}$ listed.
Column C shows the meson coupled to the $W^+$ boson.}
\label{pspspstn}
\btab{llcl}
Decay mode& C & Tree diagram of MD  & Notes\\
\hline

$D^+\ra\pi^0 K^+$ & $D^{*+}$ & $(2.9\pm1.1)\dek{-4}$ & (a)\\

$D^+\ra\eta K^+$ & $D^{*+}$ & $(5.2\pm1.7)\dek{-4}$ & (a)\\

$D^+\ra\eta^\prime K^+$ & $D^{*+}$ & $<5.7\dek{-4}$ & (a,b)\\

$\dsp\ra\eta K^+$ & $\dssp$ & $(1.8\pm0.5)\dek{-3}$ &  \\

$\dsp\ra\eta^\prime K^+$ & $\dssp$ & $(1.3\pm0.5)\dek{-3}$ &  \\

$\dsp\ra K^0 K^+$ & $D^{*+}$ & $<5.6\dek{-4}$ & (a,c,d) \\

$B^+\ra\ad K^+$ & $\bcsp$ & $(2.8\pm0.7)\dek{-4}$ &(a,e) \\

$B^+\ra\ad D^+$ & $\bcsp$ & $(4.0\pm1.4)\dek{-4}$ &(a,e,f) \\

$B^0\ra D^-K^+$ & $\bcsp$ & $(2.7\pm0.7)\dek{-4}$ & (a,e) \\

$B^0\ra D^-D^+$ & $\bcsp$ & $(3.8\pm1.3)\dek{-4}$ & (a,e,f) \\

$B_s^0\ra D_s^-K^+$ & $\bcsp$ & $(1.1\pm0.4)\dek{-3}$ &(a) \\

$B_s^0\ra D_s^-D^+$ & $\bcsp$ & $(1.5\pm0.6)\dek{-3}$ &(a,e,f) \\

$B_s^0\ra D_s^-D_s^+$ & $\bcsp$ & $(7.4\pm4.0)\%$ &  (e)

\etab
(a) Doubly Cabibbo suppressed mode.
(b) Using $B(D^+\ra\eta^\prime\pi^+)<9\dek{-3}$.
(c) Using $B(D_s^+\ra K^0\pi^+)<8\dek{-3}$.
(d) This mode is experimentally indistinguishable from
    $\dsp\ra\ak K^+$ and represents a negligible background to it.
(e) Existence of $\bcsp$ with a mass of 6.34 GeV/$c^2$ assumed.
(f) $D^+$ decay constant taken from \cite{soni}.
\et

\bt
\caption{Branching fractions of the $P_1\ra P_2+V(A)$ decay modes
in the tree level of MD and comparison with experimental
data. Column C shows the meson coupled to the $W^+$ boson.
Only results that do not contradict data are listed.}
\label{pspsvtg}
\btab{llccl}
Decay mode& C & Meson dominance result & Data & Notes\\
\hline

$\dsp\ra\eta^\prime\rho^+$ & $\dssp$ & $(11.9\pm4.6)\%$ &
$(12\pm4)\%$ &  \\

$B^+\ra\pi^0\rho^+$&$B^{*+}$&$(1.2\pm0.4)\dek{-5}$&$<7.7\dek{-5}$\\

$B^+\ra\pi^0K^{*+}$&$B^{*+}$&$(7.4\pm2.5)\dek{-7}$&$<9.9\dek{-5}$&(a)\\

$B^+\ra\pi^0D_s^{*+}$&$B^{*+}$&$(6\pm5)\dek{-5}$&$<3.3\dek{-4}$&(b)\\

$B^+\ra\pi^0a_1^+$&$B^{*+}$&$(2.0\pm0.7)\dek{-5}$&$<1.7\dek{-3}$\\

$B^0\ra\pi^-\rho^+$&$B^{*+}$&$(2.4\pm0.8)\dek{-5}$&$<8.8\dek{-5}$& (c)\\

$B^0\ra\pi^-K^{*+}$&$B^{*+}$&$(1.4\pm0.5)\dek{-6}$&$<7.2\dek{-5}$& (a)\\

$B^0\ra\pi^-D_s^{*+}$&$B^{*+}$&$(1.1\pm1.0)\dek{-4}$&$<5\dek{-4}$& (b)\\

$B^0\ra\pi^-a_1^+$&$B^{*+}$&$(3.8\pm1.3)\dek{-5}$&$<4.9\dek{-4}$& (d)\\

$B^+\ra\eta D_s^{*+}$&$B^{*+}$&$(7\pm6)\dek{-4}$&$<8\dek{-4}$&(b)\\

$B^+\ra\ad D_s^{*+}$ & $\bcsp$ & $(2.0\pm1.7)\%$ &
$(1.2\pm1.0)\%$ & (b,e) \\

$B^+\ra\ad a_1^+$ & $\bcsp$ & $(9.8\pm2.6)\dek{-3}$ &
$(5\pm4)\dek{-3}$ & (e) \\

$B^0\ra D^-\rho^+$ & $\bcsp$ & $(6.3\pm1.7)\dek{-3}$ &
$(7.8\pm1.4)\dek{-3}$ & (e) \\

$B^0\ra D^-a_1^+$ & $\bcsp$ & $(9.4\pm2.5)\dek{-3}$ &
$(6.0\pm3.3)\dek{-3}$ & (e)

\etab
(a) Doubly Cabibbo suppressed mode.
(b) Using $(w_{D_s^*}\Vcs)^2=39\pm31$, as determined from $B(B^0\ra
    D^-D_s^{*+})=(2.0\pm1.5)\%$.
(c) Experimental value includes also the $\pi^+\rho^-$ mode.
(d) Experimental value includes also the $\pi^+a_1^-$ mode.
(e) Existence of $\bcsp$ with a mass of 6.34 GeV/$c^2$ assumed.
\et

\bt
\caption{Branching fractions of the $P_1\ra P_2+V (A)$ decay modes
in the tree level of MD and comparison with experimental
data. Column C shows the meson coupled to the $W^+$ boson.
Only results that contradict existing data are listed.}
\label{pspsvtb}
\btab{llccl}
Decay mode& C & Meson dominance result & Data & Notes\\
\hline

$D^+\ra\ak\rho^+$ & $\dssp$ & $(11.7\pm0.7)\%$ &
$(6.6\pm2.5)\%$ &  \\

$D^0\ra\ak\rho^0$ & none &0 & $(1.20\pm0.17)\%$ &  \\

$D^0\ra K^-\rho^+$ & $\dssp$ & $(4.57\pm0.27)\%$ &
$(10.8\pm1.0)\%$ &  \\

$D^+\ra\ak K^{*+}$ & $\dssp$ & $(0.6\pm0.4)\%$ & $(3.0\pm1.4)\%$ &  \\

$D^0\ra K^- K^{*+}$ & $\dssp$ &$(2.20\pm0.14)\dek{-3}$  &
$(3.5\pm0.8)\dek{-3}$ &  \\

$D^+\ra\ak a_1^+$ & $\dssp$ & $(3.77\pm0.22)\% $& $(8.1\pm1.7)\%$ &  \\

$D^0\ra K^- a_1^+$ & $\dssp$ &$(1.5\pm0.1)\%$  &
$(7.3\pm1.1)\%$ &  \\

$\dsp\ra\eta\rho^+$ & $\dssp$ & $(3.3\pm0.9)\%$ &
$(10.3\pm3.2)\%$ &  \\

$B^+\ra\ad\rho^+$ & $\bcsp$ & $(0.7\pm0.2)\%$ &
$(1.34\pm0.18)\%$ & (a)

\etab
(a) Existence of $\bcsp$ with a mass of 6.34 GeV/$c^2$ assumed.
\et

\def\hs{\hspace{2cm}}
\bt
\caption{Selected $B_c^+$ decays:
MD predictions for branching ratios and their comparison with those of
various models.}
\label{comp}
\btab{lcccccc}
Branching ratio & MD & PQCD \cite{du96} & BS \cite{chang} &
ISGW \cite{lusignol}& BSW \cite{lusignol} & BSW \cite{du89}\\
\hline
$B_c^+\ra\eta_c$\ +\ $\ell^+\nu_\ell/\pi^+$  &5.0& &4.3&4.0& &4.1\\
\hs$\tau^+\nu_\tau/\pi^+$ & 1.1\\
\hs$K^+/\pi^+$ &0.075&0.068&0.078&0.074&&0.078\\
\hs$\rho^+/\pi^+$ & 2.3&3.0&2.6&2.4&&2.6\\
\hs$\ksp/\pi^+$ & 0.10&0.09&0.14&0.12&&0.14\\
\hs$a_1^+/\pi^+$ & 2.6\\
\hs$K_1^+/\pi^+$ & 0.18\\
$B_c^+\ra B_s + \ell^+\nu_\ell/\pi^+$&0.38& &0.36&0.26&0.25&0.27\\
\hs$K^+/\pi^+$ &0.064& &0.072&0.075&0.070&0.073\\
\hs$\rho^+/\pi^+$ & 0.45& &0.77&0.46&0.40&0.60
\etab
PQCD: using perturbative QCD framework proposed in \cite{brodsky};
BS: Bethe-Salpeter description of the meson wave functions and the
hadronic matrix elements;
ISGW: model of Isgur, Scora, Grinstein, and Wise \cite{isgw};
BSW: model of Wirbel, Stech, and Bauer \cite{bsw}.
\et

\bt
\caption{Branching fractions of the flavor changing ``weak neutral
current" modes  $P_1\ra P_2+\dil$ calculated in the MD approach assuming
the dominant role of the $a_1$ resonance. Column C shows the mesons
coupled to $W^+$.}
\label{neutralt}
\btab{llccc}
Decay mode& C & MD result &Other predictions& Data  \\
\hline

$K^+\ra\pi^+\die$ & $K^+,a_1^+$ &$\approx 3.1\dek{-7}$ (a)&
  & $(2.74\pm0.23)\dek{-7}$ \\

$K^+\ra\pi^+\dim$ & $K^+,a_1^+$ &
$(6.2\pm0.5)\dek{-8}$ (b)&$(6.2^{+0.8}_{-0.6})\dek{-8}$ \cite{ecker96}&
$<2.3\dek{-7}$\\

$D^+\ra\pi^+\die$ & $D^+,a_1^+$ &
$(3.9\pm0.9)\dek{-7}$ (c)&$<10^{-8}$ \cite{hewett} &$<6.6\dek{-5}$ \\

$D^+\ra\pi^+\dim$ & $D^+, a_1^+$ &
$(3.9\pm0.9)\dek{-7}$ (c)&$<10^{-8}$ \cite{hewett}&$<1.8\dek{-5}$ \\

$D_s^+\ra\pi^+\dim$ & $D_s^+,a_1^+$ &
$(1.0\pm0.5)\dek{-5}$ (d)&&$<4.3\dek{-4}$

\etab
(a) Using the $a_1\rho\pi$ coupling constant determined from
    $\Gamma_{a_1}\approx 400$~MeV.
(b) Normalized by the $K^+\ra\pi^+\die$ experimental branching fraction.
(c) Using the $a_1\rho\pi$ coupling constant determined from
    $K^+\ra\pi^+\die$ and the lattice calculation \cite{soni} result
    for the $D^+$ decay constant.
(d) Using the $a_1\rho\pi$ coupling constant determined from
    $K^+\ra\pi^+\die$ and the $\dsp$ decay constant from
    the experimental branching fraction of $\dsp\ra\mu^+\nu_\mu$.
\et

\begin{references}

\bibitem{vmdsource}
Y.~Nambu, Phys. Rev. {\bf 106}, 1366 (1957);
W.R.~Frazer and J.R.~Fulco, Phys. Rev. Lett. {\bf 2}, 365 (1959);
J.J.~Sakurai, Ann. Phys. (N.Y.) {\bf 11} (1960) 1;
M.~Gell-Mann and F.~Zachariasen, Phys. Rev. {\bf 124}, 953 (1961);
M.~Gell-Mann, Phys. Rev. {\bf 125}, 1067 (1962);
Y.~Nambu and J.J.~Sakurai, \prl{8}{79}{62}; {\bf 8}, 191(E) (1962);
M.~Gell-Mann, D.~Sharp and W.~Wagner, \prl{8}{261}{62}.

\bibitem{standard}
S.L.~Glashow, Nucl. Phys. {\bf 22}, 579 (1961);
S.~Weinberg, Phys. Rev. Lett. {\bf 19}, 1264 (1967);
A.~Salam in {\it Proc. 8th Nobel Symposium}, Aspen\H{a}sg{\aa}rden,
1968, ed. N.~Svartholm
(Almquist and Wiksell, Stockholm, 1968) p.~367.

\bibitem{gim}S.L.~Glashow, J.~Iliopoulos, and L.~Maiani,
\prd{2}{1258}{70}.

\bibitem{chounet}L.-M.~Chounet, J.-M.~Gaillard, and M.K.~Gaillard,
Phys. Rep. C {\bf 4}, 199 (1972).

\bibitem{bardin} D.Yu. Bardin and E.A. Ivanov, Fiz. Elem. Chastits
At. Yadra {\bf 7}, 286 (1978) [Sov.~J. Part. Nucl. {\bf 7}, 286 (1978)].

\bibitem{buchalla}G.~Buchalla, A.J.~Buras, and M.E.~Lautenbacher,
\rmp{68}{1125}{96}.

\bibitem{pdg}
R.M.~Barnett {\it et al.} (Particle Data Group), \prd{54}{1}{96}.

\bibitem{isgw2}D.~Scora and W.~Isgur, \prd{52}{2783}{95}.

\bibitem{du96}D.~Du, G.~Lu, and Y.~Yang, \plb{387}{187}{96}.

\bibitem{cabibbo}N.~Cabibbo, \prl{10}{531}{63}.

\bibitem{kobmask}M.~Kobayashi and T.~Maskawa, Prog. Theor. Phys.
{\bf 49}, 652 (1973).

\bibitem{class}S.~Weinberg, Phys. Rev. {\bf 112}, 1375 (1958).

\bibitem{beldjoudi}L.~Beldjoudi, Tran~N.Truong, \plb{344}{419}{95}.

\bibitem{dubnicka}M.E.~Biagini, S.~Dubni\v{c}ka, E.~Etim, and
P.~Kol\'{a}\v{r}, Nuovo Cimento A {\bf 104}, 363 (1991).

\bibitem{soni}C.W.~Bernard, J.N.~Labrenz, and A.~Soni,
\prd{49}{2536}{94}.

\bibitem{decker96}R.~Decker, M.~Finkemeier, P.~Heiliger, and
H.H~Jonsson, \zpc{70}{247}{96}.

\bibitem{fink96}M.~Finkemeier and E.~Mirkes, \zpc{69}{243}{96}.

\bibitem{davoud}H.~Davoudias and M.B.~Wise, \prd{53}{2523}{96}.

\bibitem{colangelo}G.~Colangelo, M.~Finkemeier, and R.~Urech,
\prd{54}{4403}{96}.

\bibitem{li}B.A.~Li, Phys. Rev. D (to be published) (hep-ph/9606402).

\bibitem{tsai}Y.S.~Tsai, \prd{4}{2821}{71};

\bibitem{perl}M.L.~Perl \ea, \prl{35}{1489}{75}.

\bibitem{cvc}S.S.~Gerstein and J.B.~Zeldovich, Zh. Eksp. Teor. Fiz.
{\bf 29}, 698 (1955) [Sov. Phys.--JETP {\bf 2}, 576 (1956)];
R.P.~Feynman and M.~Gell-Mann, Phys. Rev. {\bf 109}, 193 (1958).

\bibitem{weinbsum}S.~Weinberg, \prl{18}{507}{67}.

\bibitem{dmo}T.~Das, V.S.~Mathur, and S.~Okubo, \prl{18}{761}{67}.

\bibitem{okun}L.B.~Okun, {\it Leptons and Quarks} (North-Holland Pub. Co.,
Amsterdam, New York, 1982);
{\it Leptony i kvarki} (Nauka, Moskva 1981).

\bibitem{pietschm}
H.~Pietschmann, {\it Weak Interactions--Formulae, Results and
Derivations} (Springer Verlag, Wien, New York, 1983), p.~174.

\bibitem{santamar}J.H.~K\"{u}hn and A.~Santamaria, \zpc{48}{445}{90}.

\bibitem{decker}R.~Decker, \zpc{36}{487}{87}.

\bibitem{argus87}
H.~Albrecht \ea, \plb{185}{223}{87}.

\bibitem{castro}G.~L\'{o}pez Castro and D.A.~L\'{o}pez Falc\'{o}n,
\prd{54}{4400}{96}.

\bibitem{buskulic} ALEPH Collaboration, D. Buskulic \ea,
\zpc{70}{579}{96}.

\bibitem{wesszumi}J.~Wess and B.~Zumino, \plb{37}{95}{71}.

\bibitem{chiralan}R.~Fischer, J.~Wess, and F.~Wagner, \zpc{3}{313}{80};
G.~Aubrecht, N.~Chahrouri, and K.~Slanec, \prd{24}{1318}{81}.

\bibitem{kramer84}G.~Kramer and W.F.~Palmer, \zpc{25}{195}{84}.

\bibitem{leroy}C.~Leroy and J.~Pestieau, \plb{72}{398}{78}.


\bibitem{leutroos}H.~Leutwyler and M.~Roos, \zpc{25}{91}{84}.

\bibitem{gassleut}J.~Gasser and H.~Leutwyler, \npb{250}{517}{85}.

\bibitem{potent1}S.~Godfrey and N.~Isgur, \prd{32}{189}{85}.

\bibitem{potent2}E.~Eichten and F.~Feinberg, \prd{23}{2724}{81};
S.S.~Gershtein \ea, \sjnp{48}{515}{88}{327}.

\bibitem{khodruck}A.~Khodjamirian and R.~R\"{u}ckl, preprint
WUE-ITP-96-020, MPI-PhT/96-108 (Oct. 1996) [hep-ph/9610367].

\bibitem{dennery}P.~Dennery and H.~Primakoff, Phys. Rev. {\bf 131},
1334 (1963).

\bibitem{HQET}
M.B.~Voloshin and M.A.~Shifman, \sjnp{47}{801}{88}{511};
N.~Isgur and M.B.~Wise, \plb{232}{113}{89}; {\bf 237}, 527 (1990);
E.~Eichten and B.~Hill, \ibid {234}{511}{90};
H.~Georgi, \ibid{240}{447}{90};
F.~Hussain, J.G.~K\"{o}rner, K.~Schilcher, G.~Thompson, and Y.L.~Wu,
\ibid{240}{447}{90};
A.F.~Falk, H.~Georgi, B.~Grinstein, and M.B.~Wise, \npb{343}{1}{90};
J.G.~K\"{o}rner and G.~Thompson,
\plb{240}{185}{91}; M.~Neubert, \ibid{264}{455}{91}; {\bf 338}, 84
(1994); \pr{245}{259}{94}.

\bibitem{bcstar}Y.-Q. Chen, OHSTPY-HEP-T-96-029 (September 1996),
hep-ph/9610239;
C.-H.~Chang, Y.-Q.~Chen, and R.J.~Oakes, \prd{54}{4344}{96}.

\bibitem{vain76}A.I.~Vainstein, V.I.~Zakharov, L.B.~Okun, and
M.A~Shifman, \sjnp{24}{820}{76}{427}.

\bibitem{bloch}P.~Bloch \ea, \plb{56}{201}{75}.

\bibitem{ecker87}G.~Ecker, A.~Pich,~E.~de~Rafael, \npb{291}{692}{87}.

\bibitem{berg}L.~Bergstr\"{o}m and P.~Singer, \prd{43}{1568}{91}.

\bibitem{fajfer96}S.~Fajfer, \zpc{71}{307}{96}.

\bibitem{pich96}A.~Pich, {\it Rare Kaon Decays}, invited talk at the
Workshop on K Physics, Orsay, France, May 30--June 4, 1996. Available
as hep-ph/9610243.

\bibitem{ecker96}G.~Ecker, {\it Status of Chiral Perturbation Theory},
talk given at the Workshop on Heavy Quarks at Fixed Target, St. Goar,
Germany, Oct. 3-6, 1996. To appear in the Proceedings. Also available as
hep-ph/9611346.

\bibitem{alliegro}C.~Alliegro \ea, \prl{68}{278}{92}.

\bibitem{hewett}J.L.~Hewett, T.~Takeuchi, and S.~Thomas, in {\it
Electroweak Symmetry Breaking and Beyond the Standard Model},
edited by T.~Barklow, S.~Dawson, H.E.~Haber, and S.~Siegrist
(World Scientific, in press). Also available as hep-ph/9603391.

\bibitem{geng}C.Q.~Geng, I.J.~Hsu, and Y.C.~Lin, \plb{355}{569}{95}.


\bibitem{gilman}F.J.~Gilman and S.H.~Rhie, \prd{31}{1066}{85};

\bibitem{eidelman}S.I.~Eidelman and V.N.~Ivanchenko, \plb{257}{437}{91};
Nucl. Phys. B Proc. Suppl. {\bf 40}, 131 (1995).

\bibitem{cleojuly} CLEO Collaboration, J.P.~Alexander \ea,
CLNS-96-1419 (July 1996).

\bibitem{chang}C.-H.~Chang and Y.-Q.~Chen, \prd{49}{3399}{94}.

\bibitem{lusignol}M.~Lusignoli and M.~Masetti, \zpc{51}{549}{91}.

\bibitem{du89}D.~Du and Z.~Wang, \prd{39}{1342}{89}.

\bibitem{brodsky}
S.J.~Brodsky and G.P.~Lepage, \prd{22}{2157}{80};
S.J.~Brodsky and C.R.~Ji, \prl{55}{2257}{85};
A.~Szczepaniak, E.M.~Henley, and S.J.~Brodsky, \plb{243}{287}{90}.

\bibitem{isgw}
N.~Isgur, D.~Scora, B.~Grinstein, and M.~Wise, \prd{39}{799}{89}.

\bibitem{bsw}
M.~Wirbel, B.~Stech, and M.~Bauer, \zpc{29}{637}{85}.


\end{references}
\end{document}